\documentclass[aps,prd,twocolumn,superscriptaddress,eqsecnum,nofootinbib]{revtex4}

\usepackage{color}
\usepackage{amssymb}
\usepackage{amsmath}
\usepackage{amsfonts}
\usepackage{longtable}
\newcommand{\be}{\begin{equation}}
\newcommand{\ee}{\end{equation}}
\newcommand{\bea}{\begin{eqnarray}}
\newcommand{\eea}{\end{eqnarray}}
\newcommand{\bes}{\begin{subequations}}
\newcommand{\ees}{\end{subequations}}
\newcommand{\nn}{\nonumber}

\newcommand{\hn}{\hat{\bm{n}}}
\newcommand{\bv}{\bm{v}}
\newcommand{\hbv}{\hat{\bm{\lambda}}}
\newcommand{\hbn}{\hat{\bm{n}}}
\newcommand{\rd}{\dot{r}}
\newcommand{\mr}{\frac{m}{r}}
\newcommand{\mrsq}{\frac{m^2}{r^2}}
\newcommand{\mrcb}{\frac{m^3}{r^3}}
\newcommand{\hLN}{\hat{\bm{L}}_\mathrm{N}}
\newcommand{\dhLN}{\dot{\hat{\bm{L}}}_\mathrm{N}}
\newcommand{\dm}{\frac{\delta m}{m}}
\newcommand{\Sp}{\bm{S}}
\newcommand{\bc}{\bm{\chi}}

\newcommand{\D}{\bm{\Delta}}

\usepackage{bm}
\usepackage{graphicx}

\begin{document}

\title{Recoil velocity at 2PN order for spinning black hole binaries}

\author{\'{E}tienne Racine}
\author{Alessandra Buonanno}
\affiliation{Maryland Center for Fundamental Physics, Department of Physics, University of Maryland, College Park, MD 20742}
\author{Larry Kidder}
\affiliation{Center for Radiophysics and Space Research, Cornell University, Ithaca, NY 14853}
\date{\today}

%%%%%%%%%%%%%%%%%%%%%%%%%%%%%%%%%%%%%%%%%%%%%%%%
\begin{abstract}

  We compute the flux of linear momentum carried by gravitational
  waves emitted from spinning binary black holes at 2PN order for
  generic orbits. In particular we provide explicit expressions of
  three new types of terms, namely next-to-leading order spin-orbit
  terms at 1.5 PN order, spin-orbit tail terms at 2PN order, and
  spin-spin terms at 2PN order. Restricting ourselves to quasi-circular orbits,
  we integrate the linear-momentum flux over time to obtain 
  the recoil velocity as function of orbital frequency. We find 
  that in the so-called superkick configuration the higher-order 
  spin corrections can increase the recoil velocity up to a 
  factor $\sim 3$ with respect to the leading-order PN prediction. Whereas the 
  recoil velocity computed in PN theory within the adiabatic approximation 
can accurately describe the early inspiral phase, we find that its 
  fast increase during the late inspiral and plunge, 
  and the arbitrariness in determining until when it should 
  be trusted, makes the PN predictions for the total recoil 
  not very accurate and robust. Nevertheless, the linear-momentum 
  flux at higher PN orders can be employed to build more reliable 
  resummed expressions aimed at capturing the 
  non-perturbative effects until merger. Furthermore, we 
  provide expressions valid for generic orbits, and accurate at 2PN
  order, for the energy and angular momentum carried by gravitational
  waves emitted from spinning binary black holes. Specializing to
  quasi-circular orbits we compute the spin-spin terms at 2PN order in
  the expression for the evolution of the orbital frequency and found
  agreement with Mik\'{o}czi, Vas\'{u}th and Gergely. We also verified that in
  the limit of extreme mass ratio our expressions for the energy and
  angular momentum fluxes match the ones of Tagoshi, Shibata, Tanaka and Sasaki
  obtained in the context of black hole perturbation theory.

\end{abstract}
%%%%%%%%%%%%%%%%%%%%%%%%%%%%%%%%%%%%%%%%%%%%%%%%
\maketitle

\section{Introduction}

\subsection{Motivation and summary of results}

In the past few years the study of linear momentum carried by
gravitational radiation and the subsequent recoil (or kick) velocity it imparts to
a binary merger has received a lot of attention, as this recoil effect
is astrophysically very relevant~\cite{Merritt}. There has been a lot
of effort devoted in quantifying the impact of gravitational recoil on
stellar mass black hole population and supermassive black hole (SMBH) growth
scenarios~\cite{MQ,Haiman,TH,Volonteri,VP,LCFH,BL,MAS}, as well as on
formation of galactic cores~\cite{BKMQ,GM}. In addition the
observation of a candidate black hole ejected from its host galaxy
after a merger has recently been reported~\cite{KZL}. The evidence for
a recoiling black hole lies in the detection of broad blueshifted
emission lines, presumably from gas carried by the recoiling hole,
accompanied by a corresponding set of narrow emission lines from the
gas left behind in the host object. The estimated recoil velocity from the line blueshift
is $\sim 2650\, {\rm km/s}$. However Refs.~\cite{Bogdanovic,Dotti}
have proposed alternative scenarios to explain the observations of Ref.~\cite{KZL}
based on massive black hole binary
models. 

Recent estimates from binary black hole merger
simulations~\cite{Baker2,Pollney,Koppitz,jena-no-spin,Herrmann,Gonzalez,Campanelli,
recoilFAU,Bruegmann,Baker,Lousto1,Lousto2,Lousto3} 
indicate that for some special spin configurations 
recoil velocities of order $\sim 4000 \, {\rm km/s}$ could 
occur in nature. Such high kicks are easily strong enough to
eject the black hole remnant from its host galaxy. Therefore a precise
understanding of the magnitude of kick velocities and their dependence
on binary parameters is paramount for the development of accurate
galactic population synthesis models and massive black hole formation
scenarios. It is however currently impossible to simulate the number
of mergers required to span the expected binary parameter space when spins are
included, due to overwhelming computational cost. One must therefore develop analytical models 
for the recoil velocity as function of the masses and spins of the black holes. 

The first
computations of gravitational recoil in binary systems were performed
by Fitchett~\cite{Fitchett1} and later by Fitchett and
Detweiler~\cite{Fitchett2}. These papers relied upon earlier work by
Peres~\cite{Peres} and Bekenstein~\cite{Bekenstein}, who independently
computed leading-order expressions for linear momentum flux carried by
gravitational waves in terms of interference between multipole moments
of the radiation field. Fitchett's~\cite{Fitchett1} result is limited to
the regime where the binary's dynamics can be accurately described by
Newtonian physics supplemented by dissipative terms due to emission of
gravitational waves. It therefore becomes rather inaccurate when the
binary is near merger, which is where most of the recoil is
accumulated. Thus it is imperative to include post-Newtonian (PN)
corrections within this particular framework. To obtain
correct recoil velocities at higher PN order, one must use the
equations of motion for the binary at the appropriate PN order, and
also include additional couplings between multipole moments of the
radiation field. This extension of the work of Fitchett at 1PN order
for non-spinning binaries has been performed by
Wiseman~\cite{Wiseman}. More recently the computation of the recoil
for non-spinning binaries at 2PN order has been reported by Blanchet,
Qusailah and Will~\cite{BQW} [henceforth BQW].

If the binary contains spinning black holes, then the spins of the
holes contribute additional terms to the linear momentum
flux. Kidder~\cite{Kidder} has computed the leading-order (spin-orbit)
contributions from the spins to the recoil. These contributions turn
out to be 0.5PN order relative to Fitchett's calculation, i.e. to the
leading Newtonian order, showing that spins, if large, play a crucial
role in determining the recoil, especially since they introduce extra
asymmetries in the binary. In this paper we improve the
work of Kidder by computing the kick velocity including all spin
effects up to 2PN order beyond Fitchett's leading-order computation. The new
contributions we compute in this paper are the next-to-leading order
spin-orbit terms at 1.5PN order, spin-orbit tail terms at 2PN order,
and the leading spin-spin terms at 2PN order. These terms include in
particular contributions from the quadrupole moment of each spinning
black hole, which affect both the orbital equations of motion at 2PN
order, and the time evolution of the spins themselves at 1.5PN,
through precession induced by quadrupole-monopole coupling (see for
example Refs.~\cite{Racine,DamourSpinEOB}).

The PN computations just described can of course claim to provide
  a reliable estimate of the recoil velocity accumulated solely during
  the early inspiral phase preceding the plunge, merger and ringdown. However since
  a significant amount of the total recoil is generated during the
  plunge, merger and ringdown~\cite{DG,anatomy}, resummation methods 
  and the inclusion of quasi-normal modes must be invoked in
  order to provide a complete analytical model of the recoil. Preliminary 
attempts in this direction were pursued in Refs.~\cite{DG,anatomy,SB}
within the effective-one-body approach~\cite{EOB1,EOB2,EOB3}.  
Our present work pushing the calculation of the PN-expanded 
linear-momentum flux at higher orders provides a foundation for 
constructing more accurate resummed versions of the linear-momentum flux. 

Our paper is structured as follows. We first complete our introductory
section with a summary of the notation and conventions employed
throughout. Next in Sec.~\ref{sec:spins} we provide a somewhat
detailed overview of the treatment of spins in general relativity and
in PN theory. We define carefully different spin variables appearing
at various steps of our computations, e.g. spin variables of the PN
source multipole moments and spin variables with constant
magnitude. In Sec.~\ref{sec:dPdt} we outline the main computation and
give our main results for generic orbits. In Sec.~\ref{sec:circ} we
specialize our results to quasi-circular orbits and integrate the
momentum flux to obtain the kick velocity. We provide numerical
estimates of the kick velocity accumulated throughout the inspiral for
specific configurations, namely equal mass binaries with spins equal
in magnitude but opposite in direction. The spins are either collinear
with the orbital angular momentum or lying in the orbital
plane. In the collinear case we also provide an estimate of the kick
for equal masses but unequal spins. 
Finally in Sec.~\ref{sec:EJfluxes}, we provide the expressions
for the fluxes of energy and angular momentum accurate at 2PN order for
spinning binary black holes. We provide flux expressions for generic orbits 
and compare with existing results in the
literature. More specifically we verify that in the extreme mass-ratio
limit our fluxes match the formulas obtained by Tagoshi {\it et
  al.}~\cite{Tagoshi} in the framework of black hole perturbation
theory. We also compute the spin-spin terms at 2PN order in the
expression for the evolution of the orbital frequency derived from the
usual balance argument, and verify that it matches the expression
obtained by Mik\'{o}czi, Vas\'{u}th and Gergely~\cite{MVG} when one substitutes
the proper expression for the quadrupole moment of a Kerr black hole,
and one neglects contributions from magnetic dipoles.

\subsection{Conventions}

In this paper we consider black holes which can be nearly maximally
spinning. To reflect this property, our PN counting for spin
variables is defined as follows. We introduce spin variables for each
body, say $\Sp^A$, which are related to the true physical spins
$\Sp_{\rm true}^A$ by 
\be
\label{rescaledspins} \Sp^A \equiv c \Sp_{\rm true}^A.  
\ee 
This rescaling stems from the fact that for maximally
spinning compact objects, the physical spin scales as $\Sp_{\rm true}
\sim G M^2 / c$, where $M$ is the body's mass. The convention for PN
order counting thus states that the physical spin is of order
0.5PN. On the other hand the rescaled spin
variables $\Sp^A$ of Eq.~\eqref{rescaledspins} are of Newtonian order and do not
contain any hidden power of the speed of light. Of course this scaling
does not apply to slowly spinning objects, for which $\Sp_{\rm true}
\sim G M^2 v_{\rm spin}/c^2$. For such bodies the rescaled
spins~\eqref{rescaledspins} are of 0.5PN order, but such slow spins
are not targeted by this work. Throughout the body of the paper we use
geometric units $G=c=1$. However we keep track of the PN order of a
given term by assigning to it a multiplicative factor which is a power
of $1/c$. This factor should not be thought of as carrying dimensions;
it appears solely as part of our PN bookkeeping. We also denote
a term of order $n$PN, i.e. scaling as $c^{-2n}$, as being
${\cal O}(2n)$. The PN harmonic coordinates are denoted as $x^\mu =
(ct,x^i)$, latin indices being spatial. Note that the time coordinate
$x^0$ carries a PN counting factor of $c$, to track the relative
smallness of time variations compared to spatial variations in PN
theory.  In addition we denote the antisymmetric permutation
symbol by $\{\mu,\nu,\lambda,\rho\}$, with $\{0,1,2,3\} = +1$. The
Levi-Civita tensor is then 
\be 
\varepsilon^{\mu\nu\lambda\rho} \equiv \frac{1}{\sqrt{-g}}\{\mu,\nu,\lambda,\rho\}.
\ee 
The version with
indices down is given by 
\be \varepsilon_{\mu\nu\lambda\rho} = -\sqrt{-g}\, \{\mu,\nu,\lambda,\rho\}, 
\ee 
which can be derived by simply lowering the indices  with
$g_{\mu\nu}$ on the upper-index version.

\section{Spin variables in general relativity}\label{sec:spins}

In general relativity the covariant treatment of spin is somewhat
delicate. We propose to start with a discussion of a few subtleties
that one encounters when dealing with spin, in order to hopefully
sweep away from the beginning any potential confusion regarding our
analysis, and also to provide an intuitive introduction to the topic
for the reader unfamiliar with these issues. For more information and 
other recent reviews of treatment of spin in general relativity 
and PN theory, the reader may consult Refs.~\cite{WS,Kidder,FBB,
Porto,HNLSO,PortoS1S2,PortoS1S1,SpinSchafer1,SpinSchafer2,SpinSchafer3}. 

\subsection{Definitions and evolution equations}

Our discussion here relies heavily on the presentation by
Wald~\cite{Wald}, which in turn is based on the works of
Beiglb\"{o}ck~\cite{Beiglbock}, Madore~\cite{Madore} and
Dixon~\cite{Dixon}. Consider a distribution of matter described by
some stress-energy tensor $T^{\mu\nu}$. At each spacetime event inside
the body, i.e. any $x^\lambda$ such that $T^{\mu\nu}(x^\lambda) \neq
0$, one may define a spacelike surface $\Sigma(x^\lambda,n^\rho)$ over
the support of $T^{\mu\nu}$ such that it is generated by all geodesics
orthogonal to a freely specifiable timelike unit vector $n^\rho$
defined at $x^\lambda$. [This surface is well-defined as long as
its generators do not develop caustics within the matter distribution;
we assume this is the case for the purpose of our discussion.] We next
define the total momentum and spin of the matter distribution
as\footnote{Here we use Riemann normal coordinates at $x^\lambda$ to
  define the integrals.}  
\bea
p^\mu(x^\lambda,n^\rho) &\equiv& \int_{\Sigma(x^\lambda,n^\rho)} T^{\mu\nu}(y^\gamma)d\Sigma_\nu(y^\gamma), \\
S^{\mu\nu}(x^\lambda,n^\rho) &\equiv& \int_{\Sigma(x^\lambda,n^\rho)}
(y-x)^{[\mu}T^{\nu]\sigma}(y^\gamma) d\Sigma_\sigma(y^\gamma), \nn \\ 
\eea
where ${}^{[\mu \dots \nu]}$ means antisymmetrization with respect to
$\mu$ and $\nu$, $y^\gamma$ are (integration) coordinates on
$\Sigma$. It is then possible to show that at each event $x^\lambda$
there is a unique timelike unit vector $q^\rho$ collinear with
$p^\mu$, i.e.  
\be q^{[\mu}p^{\nu]}(x^\lambda,q^\rho) = 0. 
\ee 
By identifying $n^\rho = q^\rho$ one selects a preferred spacelike
surface $\Sigma(x^\lambda)$ at each event inside the body. [Henceforth
we drop the reference to the choice of normal $q^\rho$ in arguments.]
With this result in hand, one can then show that there exists a unique
timelike worldline $z^\lambda(\tau)$ such that \be\label{ssc}
p_\nu(z^\lambda)S^{\mu\nu}(z^\lambda) = 0.  
\ee
A set of kinematical constraints on $S^{\mu\nu}$ like Eq.~\eqref{ssc} is called a {\it spin supplementary condition}. The spin supplementary condition~\eqref{ssc} selects a unique worldline adapted to the matter distribution, which is called the center-of-mass worldline. Its normalized tangent vector is denoted as $u^\mu = dz^\mu / d\tau$. It is the uniqueness of the worldline definition~\eqref{ssc}, first suggested by Tulczyjew~\cite{Tulczyjew}, that makes it such a natural choice. In the literature other choices of spin supplementary conditions (and thus different definitions of center-of-mass worldlines) are sometimes employed, for example the condition $u_{\mu}S^{\mu\nu} = 0$ suggested by Pirani~\cite{Pirani}. For a recent review of various spin supplementary conditions and their link to the selection of a particular center-of-mass worldline\footnote{Intuitively, one can see that a link between the selection of a center-of-mass worldline and the definition of the spin of an object must exist, since the split of the total angular momentum into ``spin'' and ``orbital'' pieces depends explicitly on the choice of center-of-mass worldline about which the ``spin'' piece is defined, e.g. see Box 5.6 
of Ref.~\cite{MTW}.}, we refer the reader to the paper of Kyrian and Semer\'{a}k~\cite{KyrianSemerak}. In addition to fixing the worldline, the spin supplementary condition \eqref{ssc} also reduces the number of independent components of the antisymmetric tensor $S^{\mu\nu}$ from six to three, the correct number of independent spin degrees of freedom expected from Newtonian physics.    

By integrating a Taylor-expanded version of the stress-energy conservation equation $\nabla_{\mu}T^{\mu\nu} = 0$, one can show~\cite{Papapetrou} that $p^\mu$ and $S^{\mu\nu}$ obey
\bes\label{pap}
\bea
\frac{Dp^{\mu}}{D\tau} &=& -\frac{1}{2}R^{\mu}_{\,\, \nu\rho\sigma}u^{\nu}S^{\rho\sigma}, \label{DpDtau} \\
\frac{DS^{\mu\nu}}{D\tau} &=& c^2[p^\mu u^\nu - p^\nu u^\mu], \label{DSDtau}
\eea
\ees
where $R^{\mu}_{\,\,\nu\rho\sigma}$ is the Riemann tensor. While the above arguments are strictly valid for material bodies with non-vanishing stress-energy tensor, it can be shown that Eqs.~\eqref{pap} can also be applied to black holes (see for example the effective field theory treatment of Porto~\cite{Porto}). 
The spin supplementary condition~\eqref{ssc} also motivates the following definition of a spin 1-form 
\be\label{Svecdef}
S^{\mu\nu} = -\frac{1}{mc}\varepsilon^{\mu\nu\rho\sigma}p_\rho S_\sigma,
\ee
where the (conserved) mass $m$ is defined through $p^\mu p_\mu = -m^2c^2$. The conservation property of the mass so defined is ensured by system~\eqref{pap}. The inverse relation is given by
\be\label{S1formdef}
S_{\mu} = \frac{1}{2mc}\varepsilon_{\mu\nu\lambda\rho}p^\nu S^{\lambda\rho},
\ee
which is obtained by requiring $S_\mu p^\mu = 0$. Finally we note that by taking a $\tau$ derivative of Eq.~\eqref{ssc}, one can derive the following relationship between the momentum and the 4-velocity
\be\label{pofu}
(p_\lambda u^\lambda) p^\mu + m^2c^2 u^\mu = \frac{1}{2c^2}S^{\mu\nu}S^{\lambda\rho}u^{\sigma}R_{\nu\sigma\lambda\rho}.
\ee
By contracting Eq.~\eqref{pofu} with the 4-velocity, one finds
\be\label{pdotu}
p_\lambda u^\lambda = -mc \left[1 + \frac{1}{2m^2c^4}u_\mu u^\sigma S^{\mu\nu}S^{\lambda\rho}R_{\nu\sigma\lambda\rho}\right]^{1/2}.
\ee
Since $S_\mu S^{\mu\nu} = 0$, relation~\eqref{pofu} between $p^\mu$ and $u^\mu$ implies the exact equivalence between the two conditions $S_\mu p^\mu = 0$ and $S_\mu u^\mu = 0$. Thus one can view the requirement $S_\mu p^\mu = S_\mu u^\mu = 0$ as stating that in the frame instantaneously comoving with the spinning particle, the time component of the spin 1-form equals zero.

\subsection{Post-Newtonian expansion of spin evolution equation}

In this paper we specialize to systems where the PN expansion is applicable. Our next task is thus the treatment of system~\eqref{pap} in the context of PN theory. For our purposes, it is sufficient to analyze the spin evolution equation~\eqref{DSDtau}, as the momentum evolution equation has already been well studied (see e.g. Ref.~\cite{FBB}) at the order we are working at in this paper (2PN). For the remainder of this paper we use harmonic coordinates. 

First of all we rewrite Eq.~\eqref{DSDtau} into an evolution equation for $S_\mu$. By simply taking a covariant derivative of Eq.~\eqref{S1formdef}, it is easy to obtain
\bea 
\frac{DS_\mu}{D\tau} &=& \frac{1}{2mc}\varepsilon_{\mu\nu\lambda\rho}S^{\lambda\rho}\frac{Dp^\nu}{D\tau} \nn \\
&=& \frac{1}{m^3c^3}p_{[\mu}S_{\nu]}R^\nu_{\,\,\alpha\beta\gamma}u^\alpha\, \varepsilon^{\beta\gamma\lambda\rho} p_\lambda S_\rho, \label{DSDtau2}
\eea
where Eq.~\eqref{DpDtau} has been used. The next step is to expand the above evolution equation in the regime where PN gravity is valid. Since we are concerned with all contributions from spins at 2PN order in the linear momentum flux, we need to check if the right-hand side of Eq.~\eqref{DSDtau2}, which is quadratic in spin, contributes to the precession equations at the order required for our computation which is 1.5PN order. To begin with, it is clear from Eqs.~\eqref{pofu} and \eqref{pdotu} that we may replace $p^\mu$ by $mc \, u^\mu$ in Eq.~\eqref{DSDtau2}, as the corrections introduced by that substitution are well beyond the PN order that interests us here. Equation~\eqref{DSDtau2} then becomes
\bea
\frac{DS_\mu}{D\tau} &=& \frac{1}{2mc}u_{\mu}S_{\nu}R^\nu_{\,\,\alpha\beta\gamma}u^\alpha\, \varepsilon^{\beta\gamma\lambda\rho} u_\lambda S_\rho \nn \\
&&  + \,\, {\rm higher\,\,  PN \,\, corrections}. \label{DSDtau3}
\eea
We next convert proper time derivatives into coordinate time derivative to bring Eq.~\eqref{DSDtau3} closer to the usual form of the precession equations. This is accomplished using
\bea
\frac{DS_\mu}{D\tau} &=& \frac{dS_\mu}{d\tau} - \Gamma_{\mu \nu}^\lambda S_\lambda u^\nu \nn \\
&=& \frac{u^0}{c}\frac{dS_\mu}{dt} -  \Gamma_{\mu \nu}^\lambda S_\lambda u^\nu, \label{DSDtau4}
\eea
where the second line follows from the parametrization $u^\mu = u^0(1, v^i/c)$, $u^0$ being determined by normalization and $v^i = dz^i/dt$ is the coordinate velocity of the center-of-mass worldline. Note here that the condition $S_\mu u^\mu = 0$ implies 
\be
S_0 = - \frac{v^i}{c}S_i,
\ee
which shows that $S_0$ is ${\cal O}(1)$. We then have
\be\label{dSidt}
\frac{dS_i}{dt} = \frac{c}{u^0}\Gamma^\lambda_{i\nu}S_\lambda u^\nu + \frac{u_i}{2m u^0}S_{\nu}R^\nu_{\,\,\alpha\beta\gamma}u^\alpha\, \varepsilon^{\beta\gamma\lambda\rho} u_\lambda S_\rho.
\ee
As we shall see later in the paper, we require the right-hand side of Eq.~\eqref{dSidt} to be accurate to 1.5PN order, or ${\cal O}(3)$. The Riemann tensor components are at least ${\cal O}(2)$ (see for example Weinberg~\cite{Weinberg}), the spatial components of the 4-velocity $u^i$ are ${\cal O}(1)$, $u^0$ is ${\cal O}(0)$ and $S_0$ is ${\cal O}(1)$. Therefore the only possibility for the term involving the Riemann tensor to contribute in our computation is the following: the index $\nu$ is spatial, the index $\alpha$ is the time index, the index $\lambda$ is the time index, and the index $\rho$ is spatial. This implies that the indices $\beta$ and $\gamma$ must also be spatial due to the antisymmetry of the Levi-Civita tensor. Thus the potential contribution comes from the components of the Riemann tensor having the structure $R^i_{0jk}$. However a direct computation shows that these components are all ${\cal O}(3)$, which, combined with the presence of $u_i$ in front, yields a total contribution at ${\cal O}(4)$. This implies that the leading contributions of the term involving the Riemann tensor in the precession equations are ${\cal O}(4)$ and do not contribute to our computation. 

The remaining term in the right-hand side of Eq.~\eqref{dSidt} contains the well-known spin-orbit and spin-spin precession terms. 
Specializing to a binary system, we obtain that the first term on the right-hand side of Eq.~(\ref{dSidt}) gives (for body 1) 
\bea
\frac{d\Sp_1}{dt} &=& \frac{m_2}{c^2 r_{12}^2}\bigg\{-(\bm{n}_{12}\cdot\bm{v}_{12}) \, \Sp_1 - 2(\bm{v}_{12}\cdot\Sp_1) \bm{n}_{12}  \nn \\
&& + (\bm{n}_{12}\cdot\Sp_1)(\bm{v}_1 - 2\bm{v}_2)\bigg\}  - \frac{1}{c^3 r_{12}^3}\bigg\{\Sp_2\times \Sp_1 \nn \\
&& - 3(\bm{n}_{12}\cdot\Sp_2)(\bm{n}_{12}\times\Sp_1)\bigg\},
\label{dSidtexpl}
\eea
where $\bm{v}_{12} = \bm{v}_1 - \bm{v}_2$ is the relative coordinate velocity, $\bm{n}_{12}$ is the unit vector pointing from body 2 to body 1, and $r_{12}$ is the coordinate orbital separation. The dot and cross products are performed with respect to the Euclidean spatial metric. The computation presented here assumes the connection coefficients appearing in Eq.~\eqref{dSidt} are generated by the other body only. It has been shown in Refs.~\cite{Owen, FBB} that the divergent self-field terms, which arise when one uses delta-function sources in the PN field equations, do not contribute to the precession equations at the order we are concerned with here. This formally justifies our streamlined overview of the derivation of the spin evolution equations which simply ignores these self-field terms, as one does for example in Newtonian physics. Alternatively one could in principle avoid using delta-function sources and arrive at the same result from a surface integral approach, which requires knowledge of the vacuum field equations alone (see e.g. Ref.~\cite{Racine2}).The result (\ref{dSidtexpl}) was originally obtained in Ref.~\cite{BOC} as the classical limit of spinning particles in quantum field theory. 

The system of equations~\eqref{pap} applies to the so-called pole-dipole model of an astrophysical object. When dealing with systems of Kerr black holes however, system~\eqref{pap} is incomplete as one must include, in principle, the contributions from all multipole moments of the Kerr black holes. However for our computations, we only need to include the contribution of the mass quadrupole moment to the orbital equations of motion~\cite{Poisson,Gergely} and to the precession equations~\cite{Racine,DamourSpinEOB}. The precession term induced by quadrupole-monopole coupling is~\cite{Racine,Racine2}
\be
\label{dSidtQM}
\left[\frac{d\Sp_1}{dt}\right]_{QM} = \frac{3}{c^3r_{12}^3}\frac{m_2}{m_1} (\bm{n}_{12} \cdot \Sp_1)( \bm{n}_{12} \times \bm{S}_1).
\ee

\subsection{Choice of fundamental spin variable}\label{sec:spinchoice}

Based on the discussion of the previous subsection, it would seem
natural to work with the {\it covariant} spin variables $(\Sp_A)_i$, $A = 1,2$, as
they are the only quantities appearing in the precession
equations. However, in Refs.~\cite{FBB,BBF} the authors computed 
the ``source multipole moments'' in terms of 
{\it contravariant} spin variables. For a major part of this work, we stick 
with the same spin variables used in Refs.~\cite{FBB,BBF}, and 
define 
\bea 
\bar{\Sp}_A^i &=& \delta^{ij}\left(1  -
  \frac{2}{c^2}\frac{m_B}{r_{12}}\right)(\Sp_A)_j. \label{newspindef}
\eea 
Combining Eqs.~(\ref{dSidtexpl}), (\ref{dSidtQM}) and (\ref{newspindef}), we 
obtain the precession equations in the center-of-mass frame in terms of the 
barred spins. They read
\bea
\frac{d\bar{\Sp}_1}{dt} &=& \frac{m_2}{c^2r_{12}^2}\bigg\{(\bm{n}_{12}\cdot\bm{v}_{12}) \, \bar{\Sp}_1 - 2(\bm{v}_{12}\cdot\bar{\Sp}_1) \bm{n}_{12} \nn \\
&& + (\bm{n}_{12}\cdot\bar{\Sp}_1)(\bm{v}_1 - 2\bm{v}_2)\bigg\} -  \frac{1}{c^3 r_{12}^3}\bigg\{(\bar{\Sp}_2\times
\bar{\Sp}_1) \nn \\
&& -
3(\bm{n}_{12}\cdot\bar{\Sp}_2)(\bm{n}_{12}\times\bar{\Sp}_1) \nn \\
&& -
3\frac{m_2}{m_1} (\bm{n}_{12} \cdot \bar{\Sp}_1)( \bm{n}_{12} \times
\bar{\Sp}_1)\bigg\}, \label{barprecess} 
\eea 
In all other sections of the
paper, unless otherwise noted, the spin variables we use refer to the
(contravariant) barred spins defined in Eq.~\eqref{newspindef}, even though we do not
carry the bars throughout, for sake of convenience. For completeness
we also provide the evolution equations in the center-of-mass frame,
written in terms of the variables 
\bes\label{SDeltadefs} \bea
\bar{\Sp} &=& \bar{\Sp}_1 + \bar{\Sp}_2, \\
\bar{\D} &=& \frac{m}{m_2}\bar{\Sp}_2 - \frac{m}{m_1}\bar{\Sp}_1.
\eea \ees These are 
\begin{widetext}
\bes\label{SDeltaevol} \bea
\frac{d\bar{\Sp}}{dt} &=& \frac{\eta m}{c^2r^2}\Bigg\{\bigg[-4(\bv\cdot\bar{\Sp}) - 2\dm(\bv\cdot\bar{\D})\bigg]\hn + \bigg[3(\hn\cdot\bar{\Sp}) + \dm(\hn\cdot\bar{\D})\bigg]\bv + \rd\bigg[2\bar{\Sp} + \dm \bar{\D} \bigg]\Bigg\} \nn \\
&& + \frac{3\eta}{c^3r^3}\Bigg\{\bigg[4(\hn\cdot\bar{\Sp}) +
2\dm(\hn\cdot\bar{\D})\bigg](\hn\times\bar{\Sp}) +
\bigg[2\dm(\hn\cdot\bar{\Sp}) +
(1-4\eta)(\hn\cdot\bar{\D})\bigg](\hn\times\bar{\D})\Bigg\}, \eea \bea
\frac{d\bar{\D}}{dt} &=& \frac{m}{c^2r^2}\Bigg\{\bigg[-2\dm(\bv\cdot\bar{\Sp}) + (-2+4\eta)(\bv\cdot\bar{\D})\bigg]\hn + \bigg[\dm(\hn\cdot\bar{\Sp}) + (1-\eta)(\hn\cdot\bar{\D})\bigg]\bv + \rd\bigg[\dm\bar{\Sp} + (1-2\eta) \bar{\D} \bigg]\Bigg\} \nn \\
&& + \frac{1}{c^3r^3}\Bigg\{\bar{\D}\times\bar{\Sp} + 3\bigg[2\dm(\hn\cdot\bar{\Sp}) + (1-4\eta)(\hn\cdot\bar{\D})\bigg](\hn\times\bar{\Sp}) + 3(1-2\eta)\bigg[2(\hn\cdot\bar{\Sp}) + \dm(\hn\cdot\bar{\D})\bigg](\hn\times\bar{\D})\Bigg\}. \nn \\
\eea \ees 
\end{widetext}
We note that the fundamental spin variables presented here differ from the ones typically encountered in
the literature. Indeed one can check that the spin evolution
equation~\eqref{barprecess} conserves $S^\mu S_\mu$,   
but they do not conserve the magnitude of a given Kerr hole's spin 
$\bar{\bm{S}}_A^i\,\bar{\bm{S}}_A^j\,\delta_{ij}$, as defined in its local asymptotic rest frame. In the literature 
spin variables which conserve the magnitude of the Kerr hole's spin 
are usually preferred, so it is essential to relate our
spin variables to spins with constant magnitude, which we denote
$\Sp_{1,2}^{\rm c}$. In the center-of-mass frame the relation between
our spin variables and spins with constant magnitude was worked out in Refs.
~\cite{FBB,BBF}. It reads
\be \Sp_A^{\rm c} = \left(1 + \frac{m_B}{c^2 r}\right)\bar{\Sp}_A -
\frac{1}{2c^2}\left(\frac{m_B}{m}\right)^2(\bv\cdot\bar{\Sp}_A)\bv +
{\cal O}(4).  \ee 
The corresponding transformation rules for $\bar{\Sp}$ and
$\bar{\D}$ are
\begin{widetext} 
\bes \bea
\Sp^{\rm c} &=& \bar{\Sp} + \frac{\eta m}{c^2r}\bigg[2\bar{\Sp} + \dm \bar{\D}\bigg]  - \frac{\eta}{2c^2}\bigg[\bv\cdot\bar{\Sp} + \dm \bv\cdot\bar{\D}\bigg]\bv, \\
\D^{\rm c} &=& \bar{\D} + \frac{m}{c^2r}\bigg[\dm \bar{\Sp} +
(1-2\eta)\bar{\D}\bigg] - \frac{1}{2c^2}\bigg[\dm \bv\cdot\bar{\Sp} +
(1-3\eta)\bv\cdot\bar{\D}\bigg]\bv, \eea \ees 
and the evolution
equations for $\Sp^c \equiv \Sp_1^{\rm c} + \Sp_2^{\rm c}$ and
$\D^{\rm c} \equiv (m/m_2)\Sp_2^{\rm c} - (m/m_1)\Sp_1^{\rm c}$ are
\bes \bea
\frac{d\Sp^{\rm c}}{dt} &=& \frac{\eta m}{2c^2r^2}(\hn\times\bv)\times\bigg[7\Sp^{\rm c} + 3\dm \D^{\rm c}\bigg]  + \frac{3\eta}{c^3r^3}\Bigg\{\bigg[4(\hn\cdot\Sp^{\rm c}) + 2\dm(\hn\cdot\D^{\rm c})\bigg](\hn\times\Sp^{\rm c}) \nn \\
&& + \bigg[2\dm(\hn\cdot\Sp^{\rm c}) + (1-4\eta)(\hn\cdot\D^{\rm
  c})\bigg](\hn\times\D^{\rm c})\Bigg\}, \label{Sconstevol} \eea \bea
\frac{d\D^{\rm c}}{dt} &=& \frac{m}{2c^2r^2}(\hn\times\bv)\times\bigg[3\dm\Sp^{\rm c} + (3-5\eta)\D^{\rm c}\bigg]  + \frac{1}{c^3r^3}\Bigg\{\D^{\rm c}\times\Sp^{\rm c} + 3\bigg[2\dm(\hn\cdot\Sp^{\rm c}) + (1-4\eta)(\hn\cdot\D^{\rm c})\bigg](\hn\times\Sp^{\rm c}) \nn \\
&& + 3(1-2\eta)\bigg[2(\hn\cdot\Sp^{\rm c}) + \dm(\hn\cdot\D^{\rm
  c})\bigg](\hn\times\D^{\rm c})\Bigg\}. \label{deltaconstevol} \eea
\ees \end{widetext}
\section{Linear momentum flux}\label{sec:dPdt}

The vacuum spacetime surrounding a PN source of gravitational
radiation can be subdivided into three distinct regions~\cite{LB,Thorne},
each delimited by specific length scales. First we have a weak-field
near-zone, where the PN expansion is valid. The near zone field is
parametrized by {\it source multipole moments}, which encode explicit
information about the source of the radiation. The near-zone extends
out to a size $\lesssim$ a typical wavelength of radiation emitted by
the system. At the boundary of the near-zone begins the local wave
zone of the system. The local wave zone is a region of spacetime where
the effects of background spacetime on wave propagation are
negligible, i.e. one can describe the gravitational field with
linearized gravity around a flat background (asymptotic rest frame of
the source). The gravitational field of the local wave zone is
parametrized by a set of {\it radiative multipole moments}, which can
be determined in terms of the source multipole moments parametrizing
the near zone. It is during this matching procedure that tail
contributions to the radiative multipole moments of the local wave
zone can be identified~\cite{BlanchetDamour}. These tail terms contain information on
scattering of the waves off the near zone curved spacetime. Outside
the local wave zone one finds the distant wave zone, where one needs
to propagate the waves on the curved spacetime separating the source
and the observer. For example when dealing with mergers of
supermassive black holes at high redshift, cosmological effects on
wave propagation must be taken into account to model the observed
waveform.

In this paper we are concerned with the recoil imparted to the center-of-mass motion of the source due to the emission of gravitational waves. Clearly this recoil should be independent of the large scale details of the background spacetime into which the source is embedded if the size of the source is much smaller than background curvature. This is certainly the case for most localized astrophysical sources of gravitational waves like binary systems, and therefore the physics of the recoil should be entirely captured by the interplay between the near zone and the local wave zone. The linear momentum carried by the waves away from the source is essentially due to interference between different radiative multipole moments. Thorne~\cite{Thorne} gives the complete expression for the linear momentum carried by gravitational radiation as an infinite sum of couplings between different radiative multipole moments of the local wave zone.

\subsection{Fundamentals}

At 2PN order the explicit expression of the linear momentum flux of Thorne~\cite{Thorne} in terms of mass and current source multipole moments ($I_L
\equiv I_{i_1...i_l}$ and $J_L \equiv J_{i_1...i_l}$ respectively)  is 
\bea
\frac{dP_i}{dt} &=& \frac{2}{63}I^{(4)}_{ijk}I^{(3)}_{jk} + \frac{16}{45}\epsilon_{ijk}I^{(3)}_{jl}J^{(3)}_{kl} + \frac{1}{c^2}\Bigg[\frac{1}{1134}I^{(5)}_{ijkl}I^{(4)}_{jkl} \nn \\
&& + \frac{1}{126}\epsilon_{ijk}I^{(4)}_{jlm}J^{(4)}_{klm} + \frac{4}{63}J^{(4)}_{ijk}J^{(3)}_{jk}\Bigg]  \nn \\
&& + \frac{1}{c^4}\Bigg[\frac{1}{59400}I^{(6)}_{ijklm}I^{(5)}_{jklm} +
  \frac{2}{14175}\epsilon_{ijk}I^{(5)}_{jlmn}J^{(5)}_{klmn} \nn \\
  && +
  \frac{2}{945}J^{(5)}_{ijkl}J^{(4)}_{jkl}\Bigg] + \,\,{\rm tail \,\,
  terms},\label{Pflux} \eea 
where $I^{(n)}_L$ and $J^{(n)}_L$ denote the $n^{\rm th}$ time derivative of $I_L$ and $J_L$. 
The tail terms are shown explicitly below in Eqs.~\eqref{PPtail} and \eqref{SOtail}. The core of the
computation consists of evaluating the right-hand side of
Eq.~\eqref{Pflux} [and also the right-hand sides of Eqs.~\eqref{PPtail} and \eqref{SOtail} below] for a binary
system. More specifically, one needs to evaluate the time derivatives
of the source multipole moments\footnote{The specific expressions for
  all source multipole moments for binary systems at required order
  are provided in Appendix \ref{App:moments}. The moments $I_{ij}$,
  $I_{ijk}$ and $J_{ij}$ are needed to ${\cal O}(4)$ accuracy, the moments
  $I_{ijkl}$ and $J_{ijk}$ are needed to ${\cal O}(2)$ accuracy and the
  moments $I_{ijklm}$ and $J_{ijkl}$ are needed at ${\cal O}(0)$ (Newtonian)
  accuracy.}, substituting the evolution equations governing the
binary's dynamics whenever required. The multipole moments can be split into
orbital contributions (non-spinning) and contributions linear
in the spins as follows\footnote{Note that at 2PN order one should
  include the mass quadrupole moment of each Kerr black hole into
  $I_{ij}$. In Appendix \ref{App:moments} we explain why the
  individual quadrupole moments do not contribute to the linear
  momentum flux at 2PN order.}
\bes
\bea
I_L &=& \mathop{I}_\textrm{\tiny NS}\!\!{}_L + \mathop{I}_\textrm{\tiny S}\!{}_L, \\
J_L &=& \mathop{J}_\textrm{\tiny NS}\!\!{}_L + \mathop{J}_\textrm{\tiny S}\!{}_L,
\eea
\ees
where
\bes
\bea
\mathop{I}_\textrm{\tiny NS}\!\!{}_L &=& \mathop{I}_\textrm{\tiny NS}^0\!\!{}_L + \frac{1}{c^2}\,\mathop{I}_\textrm{\tiny NS}^2\!\!{}_L +  \frac{1}{c^4}\,\mathop{I}_\textrm{\tiny NS}^4\!\!{}_L + {\cal O}(5), \\
\mathop{I}_\textrm{\tiny S}\!{}_L &=& \frac{1}{c^3}\,\mathop{I}_\textrm{\tiny S}^3\!{}_L + {\cal O}(5), \\
\mathop{J}_\textrm{\tiny NS}\!\!{}_L &=& \mathop{J}_\textrm{\tiny NS}^0\!\!{}_L + \frac{1}{c^2}\,\mathop{J}_\textrm{\tiny NS}^2\!\!{}_L + \frac{1}{c^4}\,\mathop{J}_\textrm{\tiny NS}^4\!\!{}_L + {\cal O}(5), \\
\mathop{J}_\textrm{\tiny S}\!{}_L &=& \frac{1}{c}\,\mathop{J}_\textrm{\tiny S}^1\!{}_L + \frac{1}{c^3}\,\mathop{J}_\textrm{\tiny S}^3\!{}_L + {\cal O}(5).
\eea
\ees
From this decomposition it should be clear that our computation requires the orbital equations of motion at 2PN order, which include, for spinning binary black holes, spin-spin and quadrupole-monopole couplings. Also since the third time derivative of $J_{ij}$ is required at 2PN accuracy, one needs the time derivative of $\displaystyle \mathop{J}_\textrm{\tiny S}^1\!{}_{ij}$ at 1.5PN accuracy, which then implies that the spin precession equations are needed at 1.5PN accuracy. To achieve that accuracy one needs to include spin-spin and quadrupole-monopole couplings in the precession equations in addition to the leading-order spin-orbit term~\cite{Racine,DamourSpinEOB}. 

The equation of motion in harmonic coordinates for the relative orbital separation $\bm{x} = r\hn$ at 2PN order is the following~\cite{Kidder,Poisson}
\be
\bm{a} = \bm{a}_\mathrm{N} + \frac{1}{c^2}\bm{a}_\mathrm{1PN} + \frac{1}{c^3}\bm{a}_\mathrm{SO} + \frac{1}{c^4}\bm{a}_\mathrm{2PN} + \frac{1}{c^4}\bm{a}_{\mathrm{S}_1\mathrm{S}_2} + \frac{1}{c^4}\bm{a}_\mathrm{QM},
\ee
where 
\bes\label{accs}
\bea
\bm{a}_\mathrm{N} &=& -\frac{m}{r^2}\hn ,
\eea
\bea
\bm{a}_\mathrm{1PN} &=& -\frac{m}{r^2}\Bigg\{\left[(1+3\eta)v^2 - \frac{3}{2}\eta\rd^2 - 2(2+\eta)\mr \right]\hn \nn \\ && -2\rd(2-\eta)\bv\Bigg\},
\eea
\bea
\bm{a}_\mathrm{SO} &=& \frac{6}{r^3}\left[(\hn \times \bv)\cdot\left(2\Sp + \dm \D\right)\right]\hn \nn \\ && - \frac{1}{r^3}\bv \times \left(7\Sp + 3\dm \D\right) \nn \\ && + \frac{3\rd}{r^3}\hn \times \left(3\Sp + \dm \D\right),
\eea
\begin{widetext}
\bea
\bm{a}_\mathrm{2PN} &=& -\frac{m}{r^2}\left\{\left[\eta(3-4\eta)v^2\left(v^2 - \frac{3}{2}\rd^2\right) - \frac{1}{2}\eta(13-4\eta)\mr v^2 + \frac{15}{8}\eta(1-3\eta)\rd^4 - (2+25\eta + 2\eta^2)\mr \rd^2 \right.\right. \nn \\
&& \left.\left. + \frac{3}{4}(12+29\eta)\mrsq\right]\hn  - \frac{1}{2}\rd\left[\eta(15+4\eta)v^2 - 3\eta(3+2\eta)\rd^2 - (4+41\eta+8\eta^2)\mr\right]\bv \right\},
\eea
\bea
\bm{a}_{\mathrm{S}_1\mathrm{S}_2} &=& -\frac{3}{\eta m r^4}\Bigg\{\bigg[\Big (\Sp_1\cdot\Sp_2) - 5(\hn\cdot\Sp_1)(\hn\cdot\Sp_2)\bigg]\hn + (\hn\cdot\Sp_2)\Sp_1 + (\hn \cdot\Sp_1)\Sp_2 \Bigg\}, 
\eea
\bea
\bm{a}_\mathrm{QM} &=& -\frac{3}{2\eta m r^4}\Bigg\{\bigg[\frac{1}{q}\Sp_1^2 + q\Sp_2^2 - \frac{5}{q}(\hn\cdot\Sp_1)^2 - 5q(\hn\cdot\Sp_2)^2\bigg]\hn + 2\bigg[\frac{1}{q}(\hn\cdot\Sp_1)\Sp_1 + q(\hn\cdot\Sp_2)\Sp_2\bigg]\Bigg\}. \nn \\ \label{aQM}
\eea
\end{widetext}
\ees 
Above $\bv = \dot{\bm{x}}$ is the coordinate relative velocity. Equation~\eqref{aQM} follows from Ref.~\cite{Poisson}. Equations~\eqref{accs} are only valid in the center-of-mass frame of the binary. The existence of this center-of-mass frame stems from the fact that the general 2PN equations of motion for a binary system composed of spinning bodies admit two conserved spatial 3-vectors $\bm{\mathcal{K}}$ and $\bm{\mathcal{P}}$ such that the combination $\bm{\mathcal{G}} = \bm{\mathcal{K}} + \bm{\mathcal{P}}t$ can be interpreted as the coordinate location of the center-of-mass. Thus the conserved 3-vector $\bm{\mathcal{P}}$ is interpreted as the center-of-mass coordinate velocity. It is possible to show~\cite{FBB} that one can perform a (2PN accurate) Poincar\'{e} transformation, under which the 2PN equations of motion are invariant, such that in the new coordinate systems one has $\bm{\mathcal{K}}^\prime + \bm{\mathcal{P}}^\prime t = 0$. This new coordinate system is defined as the 2PN center-of-mass frame of the binary\footnote{In Ref.~\cite{FBB} the center-of-mass frame is computed taking into account effects linear in spins alone. However, since the linear momentum flux scales overall as $c^{-7}$, it should be possible to find a Poincar\'{e} transformation which takes us to the center-of-mass frame at 3PN accuracy including all spin contributions.}. Note however that if one were to include dissipative (radiation-reaction) terms to the equations of motion, one would find that $\bm{\mathcal{P}}$ is not conserved anymore, leading to a radiation-reaction induced recoil, which we compute here using instead the classic balance argument. 

Given that $m_{1,2}$ and $\Sp_{1,2}$ are the masses and spins of each black hole, the symbols appearing in Eqs.~(\ref{accs}) are defined as
\bes
\bea
m &=& m_1 + m_2, \\
\delta m &=& m_1 - m_2, \\
q &=& \frac{m_1}{m_2}, \\
\eta &=& \frac{m_1m_2}{m^2}, \\
\Sp &=&  \Sp_1 + \Sp_2, \\
\D &=&  m\left(\frac{\Sp_2}{m_2} - \frac{\Sp_1}{m_1}\right).
\eea
\ees
It is also interesting to note that by introducing the vector 
\be
\Sp_0 \equiv 2\Sp + \dm \D = \left(1 + \frac{m_2}{m_1}\right)\Sp_1 + \left(1+\frac{m_1}{m_2}\right)\Sp_2,
\ee
one may quite neatly combine $\bm{a}_{S_1S_2}$ and $\bm{a}_{QM}$ as follows
\bea
\bm{a}_\mathrm{SS} &\equiv& \bm{a}_{\mathrm{S}_1\mathrm{S}_2} + \bm{a}_\mathrm{QM} \nn \\
&=& -\frac{3}{2mr^4}\Bigg\{\Big[\Sp_0^2 - 5(\hn\cdot\Sp_0)^2\Big]\hn + 2(\hn\cdot\Sp_0)\Sp_0\Bigg\}. \nn \\ \label{aSS}
\eea
This simple expression can actually be derived easily from the spin-spin Hamiltonian for binary black holes computed by Damour~\cite{DamourSpinEOB}, which depends solely on the spin combination $\Sp_0$ (and orbital elements of course). However for objects other than the Kerr black holes of general relativity, Eq.~\eqref{aSS} does not hold since the relationship between their mass quadrupole moment and their spin is different from that of a Kerr black hole. 

The tails terms are composed of two main contributions, the non-spinning tail terms and the spin-orbit tail terms, which we denote as
\be
\left(\frac{dP_i}{dt}\right)_\mathrm{tail} = \frac{1}{c^3}\left(\frac{dP_i}{dt}\right)_\mathrm{NS \,\,tail} +  \frac{1}{c^4}\left(\frac{dP_i}{dt}\right)_\mathrm{SO \,\, tail},
\ee
where
\begin{widetext}
\bea
\left(\frac{dP_i}{dt}\right)_\mathrm{NS\,\, tail} &=& \frac{4m}{63}\left\{I^{(4)}_{ijk}(t)\int_{-\infty}^t I^{(5)}_{jk}(\tau) \left[\ln\left(\frac{t-\tau}{2b}\right) + \frac{11}{12}\right]d\tau +  I^{(3)}_{jk}(t)\int_{-\infty}^t I^{(6)}_{ijk}(\tau) \left[\ln\left(\frac{t-\tau}{2b}\right) + \frac{97}{60}\right]d\tau \right\} \nn \\
&& + \frac{32m}{45}\epsilon_{ijk}\left\{I^{(3)}_{jl}(t) \int_{-\infty}^t J^{(5)}_{kl}(\tau) \left[\ln\left(\frac{t-\tau}{2b}\right) + \frac{7}{6}\right]d\tau + J^{(3)}_{kl}(t) \int_{-\infty}^t I^{(5)}_{jl}(\tau) \left[\ln\left(\frac{t-\tau}{2b}\right) + \frac{11}{12}\right]d\tau\right\}, \nn \\ \label{PPtail}
\eea
and
\bea
\left(\frac{dP_i}{dt}\right)_\mathrm{SO\,\, tail} &=& \frac{32m}{45}\epsilon_{ijk}\left\{I^{(3)}_{jl}(t) \int_{-\infty}^t J^{(5)}_{kl}(\tau) \left[\ln\left(\frac{t-\tau}{2b}\right) + \frac{7}{6}\right]d\tau + J^{(3)}_{kl}(t) \int_{-\infty}^t I^{(5)}_{jl}(\tau) \left[\ln\left(\frac{t-\tau}{2b}\right) + \frac{11}{12}\right]d\tau\right\}. \nn \\ \label{SOtail}
\eea \end{widetext}
In Eq.~\eqref{PPtail}, one may use the multipole moments at Newtonian order only, i.e. $\displaystyle I_L \rightarrow \mathop{I}_\textrm{\tiny NS}^0\!\!{}_L $ and $\displaystyle J_L \rightarrow \mathop{J}_\textrm{\tiny NS}^0\!\!{}_L $, and the time derivatives are evaluated using the Newtonian equation of motion. In Eq.~\eqref{SOtail} however, one substitutes the 0.5PN expression for the current multipole moment $J_{kl}$, i.e. $\displaystyle J_{kl} \rightarrow c^{-1}\mathop{J}_\textrm{\tiny S}^1\!{}_{kl}$, and evaluates all time derivatives using again the Newtonian equation of motion. The spin precession equation is not needed to evaluate the spin-orbit tail terms.

\subsection{Results for generic orbits}

We find the following symbolic structure for the linear momentum flux
\bea \frac{d\bm{P}}{dt} &=&
\left(\frac{d\bm{P}}{dt}\right)_\mathrm{N} +
  \frac{1}{c}\left(\frac{d\bm{P}}{dt}\right)_\mathrm{SO} +
  \frac{1}{c^2}\left(\frac{d\bm{P}}{dt}\right)_\mathrm{1PN} \nn \\ && +
  \frac{1}{c^3}\Bigg[\left(\frac{d\bm{P}}{dt}\right)_\mathrm{NL\, SO}
 \nn 
   + \left(\frac{d\bm{P}}{dt}\right)_\mathrm{NS\,\, tail}\Bigg]  \nn \\
    && + \frac{1}{c^4}\Bigg[\left(\frac{d\bm{P}}{dt}\right)_\mathrm{2PN} + \left(\frac{d\bm{P}}{dt}\right)_\mathrm{SS}  + \left(\frac{d\bm{P}}{dt}\right)_\mathrm{SO\,\, tail}\Bigg] \nn \\
. \label{dPdtsymb} \eea The new terms that we
provide in this paper are the next-to-leading spin-orbit terms (NL SO)
at 1.5PN, the terms quadratic in spins at 2PN order and the spin-orbit
tail terms at 2PN order. We evaluate the spin-orbit tail terms
explicitly only when reducing to quasi-circular orbits in the next
section.

The expressions for each flux contribution in
Eq.~\eqref{dPdtsymb} can be quite involved, so we split each term into
components along different vectors. For example we write \be
\left(\frac{d\bm{P}}{dt}\right)_\mathrm{N} =
\left[\left(\frac{d\bm{P}}{dt}\right)^{\hn}_\mathrm{N}\right] \hn +
\left[\left(\frac{d\bm{P}}{dt}\right)^{\bv}_\mathrm{N}\right] \bv \ee and
provide explicit expressions for each coefficient to avoid very
lengthy formulas. Again we provide explicit results for the
instantaneous linear momentum flux only, and leave the evaluation of
the tail contributions when specializing to quasi-circular orbits
later on. Our results for the instantaneous flux are
\bes\label{dPdtNcoeffs} \bea
\left(\frac{d\bm{P}}{dt}\right)^{\hn}_\mathrm{N} &=& \frac{8\eta^2m^4}{105r^4}\rd\dm\left(55v^2 - 45\rd^2 + 12\mr\right), \nn \\ && \\
\left(\frac{d\bm{P}}{dt}\right)^{\bv}_\mathrm{N} &=&
-\frac{16\eta^2m^4}{105r^4}\dm\left(25v^2 - 19\rd^2 +
  4\mr\right). \nn \\ \eea \ees Equations~\eqref{dPdtNcoeffs} match the
results of Kidder~\cite{Kidder}. The leading order spin-orbit flux is
given by \bes\label{dPdtSOcoeffs} \bea
\left(\frac{d\bm{P}}{dt}\right)^{\hn\times\bv}_{\rm SO}  &=& -\frac{8\eta^2 m^3}{15 r^5}\big[3\rd (\hn \cdot \D) + 2(\bv \cdot \D)\big],\nn \\ && \\
\left(\frac{d\bm{P}}{dt}\right)^{\hn\times\D}_{\rm SO}  &=& -\frac{16\eta^2 m^3}{15 r^5}v^2, \\
\left(\frac{d\bm{P}}{dt}\right)^{\bv\times\D}_{\rm SO} &=& \frac{32\eta^2
  m^3}{15 r^5}\rd.  \eea \ees Again Eqs.~\eqref{dPdtSOcoeffs} match
the expression of Kidder~\cite{Kidder}. The first PN corrections to
the flux are \begin{widetext} \bes\label{dPdt1PNcoeffs} \bea
\left(\frac{d\bm{P}}{dt}\right)^{\hn}_\mathrm{1PN} &=& \frac{4\eta^2 m^4}{945 r^4}\rd \dm  \bigg[6(851 - 779\eta)v^4 - 6(2834 - 1877\eta)\rd^2 v^2 - 3(4385 - 956\eta)\mr v^2 + 6(1843 - 1036\eta)\rd^4 \nn \\
&& + (12301 - 1168\eta)\mr \rd^2  -6(295-2\eta)\mrsq \bigg] , \\
\left(\frac{dP}{dt}\right)^{\bv}_\mathrm{1PN} &=& \frac{4\eta^2 m^4}{945 r^4}\dm\bigg[-111(25-28\eta)v^4 + 30(392-257\eta)v^2\rd^2 + 9(907 - 162\eta)\mr v^2 - 3(2663 - 1394\eta)\rd^4 \nn \\
&& - 3(2699 + 10\eta)\mr \rd^2 + 8(189 + 17\eta)\mrsq \bigg] .  \eea
\ees  \end{widetext} As mentioned in the introduction Wiseman~\cite{Wiseman}
originally computed the 1PN linear momentum flux, but his results are
presented in a format which makes is quite complicated to compare with
our expression, and so we did not perform that check. We next have the
NL SO flux, given by \begin{widetext}
\bes\label{dPdtNLSOcoeffs} \bea
\left(\frac{d\bm{P}}{dt}\right)^{\hn}_\mathrm{NL\, SO}  &=& -\frac{4\eta^2m^3}{315 r^5}\rd \Bigg\{\Big[226(14-53\eta)v^2  - 6(522-1973\eta)\rd^2+ (503 - 1817\eta)\mr \Big][\D\cdot(\hn\times\bv)] \nn \\
&& + 4\dm\left(1466v^2 - 1497\rd^2 + 265
  \mr\right)[\Sp\cdot(\hn\times\bv)] \Bigg\}\,, \eea
\bea
\left(\frac{d\bm{P}}{dt}\right)^{\bv}_\mathrm{NL\, SO}  &=&  \frac{4\eta^2m^3}{945 r^5}\Bigg\{\Big[(3968 - 15017\eta)v^2 - 3(1274-4787\eta)\rd^2  + (697 - 2503\eta)\mr\Big][\D\cdot(\hn\times\bv)] \nn \\
&& + \dm\left(7985v^2 - 8043\rd^2 + 2176 \mr\right)[\Sp\cdot(\hn\times\bv)] \Bigg\}\,,
\eea
\bea
\left(\frac{d\bm{P}}{dt}\right)^{\hn\times\bv}_\mathrm{NL \,SO}  &=& \frac{4\eta^2m^3}{945 r^5}\Bigg\{3\rd\left[4(431 - 2954\eta)v^2 -6(348 - 2057\eta)\rd^2 + (1508 + 1225\eta)\mr\right](\hn\cdot\D)  \nn \\
&& + 12\rd\dm\left[1628v^2 - 1821\rd^2 - 212\mr\right](\hn\cdot\Sp) + \Big[(-1772 + 4865\eta)v^2 + 3(1490 - 3041\eta)\rd^2 \nn \\
&& + (320 - 143\eta)\mr\Big](\bv\cdot\D) - 2\dm\left[1531v^2 -2523\rd^2 -10\mr \right](\bv\cdot\Sp)\Bigg\}\,,
\eea
\bea 
\left(\frac{d\bm{P}}{dt}\right)^{\hn\times\D}_\mathrm{NL\, SO} &=& \frac{4\eta^2m^3}{945 r^5}\Bigg\{2(661 - 484\eta)v^4 - 3(3385 - 3094\eta)\rd^2v^2  + 5(233 - 596\eta)\mr v^2  + 9(1143-1090\eta)\rd^4 \nn \\
&& + 3(1195 + 1502\eta)\mr \rd^2\Bigg\} \,,
\eea
\bea
\left(\frac{d\bm{P}}{dt}\right)^{\hn\times\Sp}_\mathrm{NL\, SO} &=& \frac{4\eta^2m^3}{945 r^5}\dm\Bigg\{1469 v^4 - 7491\rd^2v^2 + 412 \mr v^2  + 6300\rd^4 - 1812\mr \rd^2 - 72 \mrsq \Bigg\} \,,
\eea
\bea
\left(\frac{d\bm{P}}{dt}\right)^{\bv\times\D}_\mathrm{NL\, SO} &=& \frac{4\eta^2m^3}{945 r^5}\rd \Bigg\{(5677-15868\eta)v^2 - 3(2785 - 6016\eta)\rd^2 - 206(29 + 4\eta)\mr \Bigg\}\,,
\eea
\bea
\left(\frac{d\bm{P}}{dt}\right)^{\bv\times\Sp}_\mathrm{NL\, SO} &=& \frac{4\eta^2m^3}{945 r^5}\rd\dm\Bigg\{5431v^2 - 5709\rd^2 + 1112\mr\Bigg\}\,.
\eea
\ees \end{widetext}
These next-to-leading order spin-orbit contributions to the linear momentum flux are new. 
The contributions at 2PN order contain terms independent of the spins and terms quadratic in spins. We first give the non-spinning terms \begin{widetext}
\bes\label{dPdt2PNNScoeffs}
\bea
\left(\frac{d\bm{P}}{dt}\right)^{\hn}_\mathrm{2PN} &=& \frac{2\eta^2 m^4}{10395r^4}\dm\rd \Bigg\{3(2040 - 187945\eta + 149936\eta^2)v^6 + 3(42464 + 900359\eta - 503040\eta^2)\rd^2 v^4 \nn \\
&& - 3(229227 - 458683\eta + 178873\eta^2)\mr v^4 - 24(5363 + 150719\eta - 65604\eta^2)\rd^4v^2 \nn \\
&& + (2634273 - 4982252\eta + 1391403\eta^2)\mr \rd^2 v^2 + (1515304-754361\eta + 212216\eta^2)\mrsq v^2 \nn \\
&& + 60(84 + 24713\eta - 8792\eta^2)\rd^6 - 3(658810 - 1128391\eta + 259236\eta^2)\mr \rd^4 \nn \\
&& - (1606846 - 562815\eta + 86622\eta^2)\mrsq \rd^2 + 2(52781 + 94638\eta - 3642\eta^2)\mrcb\Bigg\}\,, \eea
\bea
\left(\frac{d\bm{P}}{dt}\right)^{\bv}_\mathrm{2PN} &=& \frac{2\eta^2 m^4}{31185 r^4}\dm\Bigg\{-18(15482 - 54215\eta + 45928\eta^2)v^6 + 18(73439 - 307240\eta + 153180\eta^2)\rd^2v^4 \nn \\
&& + 9(141321 - 214813\eta + 80173\eta^2)\mr v^4 -18(121084 - 429935\eta + 153642\eta^2)\rd^4 v^2 \nn \\
&&  - 9(599979 - 979482\eta + 221851\eta^2)\mr \rd^2 v^2 - 2(955835 - 265551\eta + 89829\eta^2) \mrsq v^2 \nn \\
&& + 18(61339 - 177850\eta + 48782\eta^2)\rd^6 + 3(1448844 - 2083359\eta + 336232\eta^2)\mr \rd^4 \nn \\
&&  + 6(381131 - 62105\eta - 10855\eta^2)\mrsq \rd^2  -(472694 + 413208\eta - 26472\eta^2)\mrcb \Bigg\} .
\eea
\ees \end{widetext}
As far as we are aware, expression~\eqref{dPdt2PNNScoeffs} for the 2PN non-spinning linear momentum flux for generic orbits has not been reported before. When specialized to quasi-circular orbits, expression~\eqref{dPdt2PNNScoeffs} matches the one of BQW. Finally we present the 2PN contributions to the linear momentum flux that are quadratic in spins. These terms have never been computed before, and are given by
\bes\label{dPdt2PNSScoeffs} \begin{widetext}
\bea
\left(\frac{d\bm{P}}{dt}\right)^{\hn\times\bv}_\mathrm{SS} &=& \frac{4\eta^2m^2}{45r^6}\Bigg\{6\bigg[\dm\big[\rd(85+98\eta)(\hn\cdot\D)  + (13-16\eta)(\bv\cdot\D)\big] \nn \\
&& + 2\rd(37+98\eta)(\hn\cdot\Sp) - 2(11+16\eta)(\bv\cdot\Sp)\bigg]\big[(\hn\times\bv)\cdot\D\big]\nn \\
&& + 6\bigg[\rd(47+196\eta)(\hn\cdot\D)  + 8(7-4\eta)(\bv\cdot\D) + 2\dm\big[-49\rd(\hn\cdot\Sp) + 8(\bv\cdot\Sp)\big]\bigg]\big[(\hn\times\bv)\cdot\Sp\big] \nn \\
&& -\bigg(3v^2 - 150\rd^2 + 4\mr\bigg)\big[(\Sp\times\D)\cdot\hn\big] + 3\rd\big[(\Sp\times\D)\cdot\bv\big] \Bigg\},
\eea 
\bea
\left(\frac{d\bm{P}}{dt}\right)^{\D}_\mathrm{SS} &=& \frac{4\eta^2m^2}{315r^6}\Bigg\{\rd\dm\left[3(601+404\eta)v^2 - 3(615 + 548\eta)\rd^2 + 4(31-180\eta)\mr\right](\hn\cdot\D) \nn \\
&& - \dm\left[3(463+4\eta)v^2 -3(389+88\eta)\rd^2 + 8(5-18\eta)\mr\right](\bv\cdot\D) \nn \\
&& + 2\rd\left[6(347 + 202\eta)v^2 - 3(849 + 548\eta)\rd^2 + (187-720\eta)\mr\right](\hn\cdot\Sp) \nn \\
&& - 2\left[12(62+\eta)v^2 -12(95+22\eta)\rd^2 + (37-144\eta)\mr\right](\bv\cdot\Sp)\Bigg\},
\eea
\bea
\left(\frac{d\bm{P}}{dt}\right)^{\hn}_\mathrm{SS} &=& \frac{4\eta^2m^2}{315r^6}\Bigg\{ \rd\dm\bigg[81(17-72\eta)v^2 - 3(523-2016\eta)\rd^2 + 4(53-198\eta)\mr\bigg]\D^2 \nn \\
&& + \rd\bigg[9(625 - 2592\eta)v^2 - 3(2077 - 8064\eta)\rd^2 + 2(361 - 1584\eta)\mr\bigg](\Sp\cdot\D) \nn \\ 
&& + 72\rd\dm\bigg[81v^2 - 84\rd^2 + 11\mr\bigg]\Sp^2  \nn \\
&& - 3\rd\dm\bigg[(2773-8884\eta)v^2 - 15(233-840\eta)\rd^2 + 6(45-188\eta)\mr\bigg](\hn\cdot\D)^2 \nn \\
&& + \dm\bigg[3(609-2216\eta)v^2 - 12(319 - 1775\eta)\rd^2 + 32(2-9\eta)\mr\bigg](\hn\cdot\D)(\bv\cdot\D) \nn \\
&& -3\rd\dm(87+1240\eta)(\bv\cdot\D)^2 + 3\rd(423-4960\eta)(\bv\cdot\Sp)(\bv\cdot\D)  + 3720\rd\dm(\bv\cdot\Sp)^2 \nn \\
&& -3\rd\bigg[(10087 - 35536\eta)v^2 - 15(919 - 3360\eta)\rd^2 + 6(179-752\eta)\mr\bigg](\hn\cdot\D)(\hn\cdot\Sp) \nn \\
&& + \bigg[3(1545-4432\eta)v^2 - 3(4361 - 14200\eta)\rd^2 +(14-576\eta)\mr\bigg](\hn\cdot\D)(\bv\cdot\Sp) \nn \\ 
&& + \bigg[24(117 - 554\eta)v^2 - 3(2471 - 14200\eta)\rd^2 - 4(7+ 144\eta)\mr\bigg](\hn\cdot\Sp)(\bv\cdot\D) \nn \\
&& -12\rd\dm\bigg[2221v^2 - 3150\rd^2 + 282\mr\bigg](\hn\cdot\Sp)^2 + 12\dm\bigg[554v^2 - 1775\rd^2 + 24\mr\bigg](\hn\cdot\Sp)(\bv\cdot\Sp) \Bigg\},\nn \\ && \, 
\eea
\bea
\left(\frac{d\bm{P}}{dt}\right)^{\bv}_\mathrm{SS} &=& \frac{4\eta^2m^2}{315r^6}\Bigg\{ -\dm\bigg[9(81-416\eta)v^2 - 9(107-448\eta)\rd^2 + 2(91-360\eta)\mr\bigg]\D^2 \nn \\
&& - \bigg[9(365 - 1664\eta)v^2 - 9(443 - 1792\eta)\rd^2 + 2(361 - 1440\eta)\mr\bigg](\Sp\cdot\D) \nn \\
&& - 144\dm\bigg[26v^2 - 28\rd^2 + 5\mr\bigg]\Sp^2 \nn\\
&& + \dm\bigg[3(1223-4404\eta)v^2 - 3(2043-7616\eta)\rd^2 + 4(143-576\eta)\mr\bigg](\hn\cdot\D)^2 \nn \\
&& +51\rd\dm(53-216\eta)(\hn\cdot\D)(\bv\cdot\D) -3\dm(3-664\eta)(\bv\cdot\D)^2 \nn \\
&& +4\bigg[3(1163 - 4404\eta)v^2 - 12(507 - 1904\eta)\rd^2 + (575-2304\eta)\mr\bigg](\hn\cdot\D)(\hn\cdot\Sp) \nn \\
&& + 6\rd(1171-3672\eta)(\hn\cdot\D)(\bv\cdot\Sp)  + 3\rd(1607-7344\eta)(\hn\cdot\Sp)(\bv\cdot\D) -3(333-2656\eta)(\bv\cdot\Sp)(\bv\cdot\D) \nn \\
&& +12\dm\bigg[1101v^2 - 1904\rd^2 + 192\mr\bigg](\hn\cdot\Sp)^2 + 11016\rd\dm(\hn\cdot\Sp)(\bv\cdot\Sp) -1992\dm(\bv\cdot\Sp)^2\Bigg\}, \nn \\
\eea
\bea
\left(\frac{d\bm{P}}{dt}\right)^{\Sp}_\mathrm{SS} &=& \frac{4\eta^2m^2}{315r^6}\Bigg\{\rd\left[-3(523 - 808\eta)v^2 + 3(843 - 1096\eta)\rd^2 + 2(187-720\eta)\mr\right](\hn\cdot\D) \nn \\
&& + 2\left[-3(255 + 4\eta)v^2 + 3(107+88\eta)\rd^2 + (47+144\eta)\mr\right](\bv\cdot\D) \nn \\
&& -12\rd\dm\left[101v^2 - 137\rd^2 - 60\mr\right](\hn\cdot\Sp) + 12\dm\left[v^2 -22\rd^2 - 12\mr\right](\bv\cdot\Sp)\Bigg\}, 
\eea
\bea
\left(\frac{d\bm{P}}{dt}\right)^{\hn\times\D}_\mathrm{SS} &=& \frac{4\eta^2m^2}{45r^6}\Bigg\{7\left(21v^2 + 39\rd^2 + 4\mr \right)[\Sp\cdot(\hn\times\bv)] + \dm\left(93v^2 + 93\rd^2 - 2\mr \right)[\D\cdot(\hn\times\bv)]\Bigg\}, \nn \\
&& 
\eea\end{widetext}
\bea \left(\frac{d\bm{P}}{dt}\right)^{\bv\times\D}_\mathrm{SS} &=&
-\frac{8\eta^2m^2}{15r^6}\rd \Bigg[65\Sp +
38\dm\D\Bigg]\cdot(\hn\times\bv)\,.  \nn \\
\eea \ees 
This concludes the presentation of our results for generic orbits. We
remind the reader that the spins appearing in all the formulas of this
section are the (contravariant) barred spins of Sec.~\ref{sec:spinchoice}. Since the
difference between the barred spins of Sec.~\ref{sec:spinchoice}
and the more often encountered spins with constant magnitude is at 1PN
order, i.e. an ${\cal O}(2)$ difference, and linear in the spins, only
the next-lo-leading order spin-orbit linear momentum flux components
given by Eqs.~\eqref{dPdtNLSOcoeffs} are affected by this change of
variables. 

\section{Reduction to approximate circular orbits}\label{sec:circ}

In binary systems where spins are dynamically negligible, it is
well-known that emission of gravitational radiation pushes the
eccentricity of the instantaneous osculating orbit toward zero. In PN
theory of point-particles this osculating orbit is simply found by
setting $\rd = 0$ and solving the resulting equation of motion for the
angular frequency, which leads to the familiar PN generalized Kepler's
law. When spins are present however, exact circular motion is not a
solution to the equations of motion generically\footnote{The exception
  is when spins are collinear with the orbital angular momentum.}. But
since the spin-orbit and spin-spin accelerations terms responsible for
the absence of exact circular motion are of 1.5PN and 2PN order
respectively, it is still expected that the instantaneous osculating
orbit of a black hole binary should be nearly circular when entering
the LIGO band. 

Following Poisson~\cite{Poisson}, we describe this
nearly circular motion by treating the spin-dependent acceleration
terms as a perturbation, and linearize about circular motion. This
procedure is straightforward and leads to the following time-dependent expressions
for the orbital separation and frequency, which are derived in Appendix \ref{CircRedux}
\bes\label{QCorbit}
\bea
r(t) %&=& \bar{r} + \frac{1}{m^2\bar{r}}\Bigg[\frac{1}{2}(\hn \cdot\Sp_0^{\rm c})^2 + (\hbv \cdot\Sp_0^{\rm c})^2 \nn \\
%&& -\frac{3}{4}\big[(\Sp_0^{\rm c})^2 - (\hLN \cdot\Sp_0^{\rm c})^2\big]\Bigg] \nn \\
&=& \bar{r} + \frac{1}{4m^2\bar{r}}\big[(\hbv \cdot\Sp_0^{\rm c})^2 - (\hn \cdot\Sp_0^{\rm c})^2\big],\label{QCorbit1} \\
\omega(t) %&=& \bar{\omega} + \frac{\bar{\omega}}{m^2\bar{r}^2}\Bigg[-\frac{5}{2}(\hn\cdot\Sp_0^{\rm c})^2 - 2(\hbv \cdot\Sp_0^{\rm c})^2 \nn \\
%&& + \frac{9}{4}\big[(\Sp_0^{\rm c})^2 - (\hLN \cdot\Sp_0^{\rm c})^2\big]\Bigg]\nn \\
&=& \bar{\omega} + \frac{\bar{\omega}}{4m^2\bar{r}^2}\big[(\hbv \cdot\Sp_0^{\rm c})^2 - (\hn \cdot\Sp_0^{\rm c})^2\big].\label{QCorbit2}
\eea
\ees
Above $\bar{r}$ and $\bar{\omega}$ are the orbital averages of $r(t)$ and $\omega(t)$, and the vector $\hbv$ is given by 
\be
\hbv = \hLN \times \hn,
\ee 
with $\hLN = (\hn \times \bv) / |\hn \times \bv|$, so that $\hn, \,\hbv$ and $\hLN$ form a right-handed orthonormal basis. In this section we express all our results in terms of spin
variables with constant magnitudes~\cite{FBB,BBF}. The averages $\bar{r}$ and $\bar{\omega}$ are related by a modified version of Kepler's law given by 
\bea(m\bar{\omega})^2 &=& \frac{m^3}{\bar{r}^3}\Bigg\{1 -
(3-\eta)\frac{m}{\bar{r}} + \bigg(6 + \frac{41}{4}\eta + \eta^2\bigg)\frac{m^2}{\bar{r}^2} \nn \\
&& - \left[\left(5\frac{\Sp^{\rm c}}{m^2} + 3\dm
    \frac{\D^{\rm
        c}}{m^2}\right)\cdot\hLN\right]\frac{m^{3/2}}{\bar{r}^{3/2}} \nn \\
        && - \frac{3}{4m^4}\Big[(\Sp_0^{\rm
  c})^2 - 3(\Sp_0^{\rm c}\cdot\hLN)^2\Big]\frac{m^2}{\bar{r}^2}\Bigg\}, \label{modKepler} 
\eea
Relation~\eqref{modKepler} can be inverted to provide the ratio $m/\bar{r}$ 
as 
\bea
\frac{m}{\bar{r}} &=&  x\Bigg\{1 + \left(1 -
  \frac{\eta}{3}\right)x + \bigg(1 -
\frac{65}{12}\eta\bigg)x^2 \nn \\ &&  +
\left[\left(\frac{5}{3}\frac{\Sp^{\rm c}}{m^2} + \dm \frac{\D^{\rm
        c}}{m^2}\right)\cdot\hLN\right]x^{3/2} \nn \\ && +  \frac{1}{4m^4}\big[(\Sp_0^{\rm c})^2 -
3(\Sp_0^{\rm c}\cdot\hLN)^2\big]x^2\Bigg\},  \label{rofomega}
\eea
where $x = (m\bar{\omega})^{2/3}$ following BQW. Lastly the orbital velocity is expressed as 
\be
\label{vcirc}
\bv = \rd(t) \hn + \omega(t) r(t) \, \hbv.
\ee
At 2PN accuracy we can drop $\rd^2$ as it is a 4PN quantity, i.e. $\mathcal{O}(8)$, and we have
\be\label{vcirc2}
v^2 = \omega^2(t) r^2(t).
\ee

\subsection{Instantaneous linear momentum flux}

To obtain the linear momentum flux in the limit of quasi-circular orbits , one substitutes Eqs.~\eqref{QCorbit}, Eq.~\eqref{vcirc}, and then the PN expansion~\eqref{rofomega} 
into Eqs.~\eqref{dPdtNcoeffs}, \eqref{dPdtSOcoeffs}, \eqref{dPdt1PNcoeffs}, \eqref{dPdtNLSOcoeffs}, \eqref{dPdt2PNNScoeffs} and \eqref{dPdt2PNSScoeffs}. The non-spinning contribution is found to be
\bea
\left(\frac{d\bm{P}}{dt}\right)_\mathrm{NS} &=& \dot{P}_\mathrm{N}\dm\Bigg\{1 - \left(\frac{452}{87} + \frac{1139}{522}\eta\right)x + \nn \\ && \left(-\frac{71345}{22968} + \frac{36761}{2088}\eta + \frac{147101}{68904}\eta^2\right)x^2\Bigg\}\hbv, \nn \\ \label{dPdtCircPP}
\eea 
where
\be\label{ConstdotPNdef}
\dot{P}_\mathrm{N} = -\frac{464\eta^2 x^{11/2}}{105}.
\ee
Equation~\eqref{dPdtCircPP} matches the instantaneous flux of BQW. This provides a good consistency check of our computations. The contributions to the linear momentum flux depending on the spins are \begin{widetext}
\bes
\bea
\left(\frac{d\bm{P}}{dt}\right)_{\rm S}^{\hn} &=&  \dot{P}_N\frac{x^2}{m^4}\Bigg\{-\frac{549}{29}\dm (\hbv\cdot\Sp^{\rm c})(\hn\cdot\Sp^{\rm c}) + \left(-\frac{103}{29} + \frac{1098}{29}\eta\right)(\hbv\cdot\D^{\rm c})(\hn\cdot\Sp^{\rm c})  \nn \\
&& + \left(-\frac{993}{116} + \frac{1098}{29}\eta\right)(\hbv\cdot\Sp^{\rm c})(\hn\cdot\D^{\rm c}) + \dm\left(-\frac{34}{29} + \frac{549}{29}\eta\right)(\hbv\cdot\D^{\rm c})(\hn\cdot\D^{\rm c})\Bigg\}, 
\eea
\bea
\left(\frac{d\bm{P}}{dt}\right)_{\rm S}^{\hbv} &=& \dot{P}_N\frac{x^{1/2}}{m^2}\Bigg\{-\frac{7}{29}(\hLN\cdot\D^{\rm c}) + \Bigg[-\dm\frac{470}{87} (\hLN\cdot\Sp^{\rm c})  + \bigg(- \frac{67}{58} + \frac{206}{29}\eta\bigg)(\hLN\cdot\D^{\rm c})\Bigg]x \nn \\
&& + \Bigg[ 10\dm (\hLN\cdot\Sp^{\rm c})^2 + (11-40\eta)(\hLN\cdot\Sp^{\rm c})(\hLN\cdot\D^{\rm c}) + \dm\left(\frac{175}{58} - 10\eta\right)(\hLN\cdot\D^{\rm c})^2 \nn \\
&& + \frac{628}{29}\dm(\hbv\cdot\Sp^{\rm c})^2 + \left(\frac{746}{29} - \frac{2512}{29}\eta\right)(\hbv\cdot\Sp^{\rm c})(\hbv\cdot\D^{\rm c}) + \dm\left(\frac{431}{58} - \frac{628}{29}\eta\right)(\hbv\cdot\D^{\rm c})^2 \nn \\
&& -\frac{918}{29}\dm (\hn\cdot\Sp^{\rm c})^2 + \left(-\frac{4071}{116} + \frac{3672}{29}\eta\right)(\hn\cdot\Sp^{\rm c})(\hn\cdot\D^{\rm c}) + \dm\left(-\frac{1107}{116} + \frac{918}{29}\eta\right)(\hn\cdot\D^{\rm c})^2\Bigg]\frac{x^{3/2}}{m^2}\Bigg\}, \nn \\
\label{PSlambda}
\eea
\bea
\left(\frac{d\bm{P}}{dt}\right)_{\rm S}^{\hLN} &=& \dot{P}_N\frac{x^{1/2}}{m^2}\Bigg\{\frac{14}{29}(\hbv\cdot\D^{\rm c}) + \Bigg[\frac{109}{116}\dm(\hbv\cdot\Sp^{\rm c}) + \left(\frac{25}{116} - \frac{57}{58}\eta\right)(\hbv\cdot\D^{\rm c})\Bigg]x \nn \\
&&  + \Bigg[ -\frac{45}{29}\dm (\hLN\cdot\Sp^{\rm c})(\hbv\cdot\Sp^{\rm c}) + \left(\frac{845}{116} + \frac{90}{29}\eta\right)(\hbv\cdot\Sp^{\rm c})(\hLN\cdot\D^{\rm c})  \nn \\
&& + \left(-\frac{225}{58} + \frac{90}{29}\eta\right)(\hbv\cdot\D^{\rm c})(\hLN\cdot\Sp^{\rm c}) + \dm\left(\frac{125}{58} + \frac{45}{29}\eta\right)(\hbv\cdot\D^{\rm c})(\hLN\cdot\D^{\rm c})\Bigg]\frac{x^{3/2}}{m^2}\Bigg\}, \nn \\
\label{PSLN}
\eea
\ees \end{widetext}
where we have projected the remaining components along $\Sp^{\rm c}$, $\D^{\rm c}$, $\hn\times\Sp^{\rm c}$ and $\hn\times \D^{\rm c}$ on the orthonormal basis formed by $\hn$, $\hbv$ and $\hLN$. This concludes our discussion of the instantaneous linear momentum flux in the limit of quasi-circular orbits.

\subsection{Tail contributions to the linear momentum flux}

The tail contributions to the linear momentum flux are formally given by Eqs. (\ref{PPtail}) and (\ref{SOtail}). Since the tails contribute at 1.5PN order (non-spinning terms) and at 2PN order (spin-orbit terms), the precession dynamics can be dropped for the purpose of computing these terms. One can easily see this as follows. The orbital plane precession originates from the spin-orbit acceleration $\bm{a}_{SO}$, and therefore introduces 1.5PN relative corrections to the non-spinning tails. These corrections thus contribute at 3PN in the flux. Similarly the spin precession dynamics introduce 1PN relative corrections to the spin-orbit tail, and thus also show up at 3PN in the flux. For quasi-circular orbits, the fact that we can ignore precession effects when computing tails implies that we can parametrize the unit vectors $\hn$ and $\hbv$ on a convenient time-independent triad as
\bes\label{nvparam}
\bea
\hn(t) &=& (\cos\phi(t),\sin\phi(t),0),\\
\hbv(t) &=& (-\sin\phi(t),\cos\phi(t),0),
\eea
\ees
where $\phi(t)$ is the orbital phase as function of time. For the purpose of evaluating the tail integrals (\ref{PPtail}) and (\ref{SOtail}), it is useful to express the unit vectors $\hn(\tau)$ and $\hbv(\tau)$, which depend on the integration variable $\tau$, as linear combinations of $\hn(t)$ and $\hbv(t)$. Doing so allows one to pull vector quantities outside of the integral over $\tau$. These linear combinations are
\bes
\bea
\hn(\tau) &=& \cos[\phi(t)-\phi(\tau)] \hn(t) - \sin[\phi(t) - \phi(\tau)] \hbv(t), \nn \\ && \\
\hbv(\tau) &=& \sin[\phi(t) - \phi(\tau)] \hn(t) + \cos[\phi(t)-\phi(\tau)] \hbv(t). \nn \\
\eea
\ees
In evaluating the time derivatives of the source multipole moments appearing in (\ref{PPtail}) and (\ref{SOtail}), it is sufficiently accurate to substitute the Newtonian equations of motion when necessary. The non-spinning tail contributions at 1.5PN order have been reported by BQW, but we review in detail their computation for sake of completeness, and also as a methodology check for our computation of the new tail terms involving the spins. 
We begin with the 1.5PN non-spinning tail terms. As a first step, one must first compute the index contractions appearing in Eq.~\eqref{PPtail}. Defining $\varphi \equiv \phi(t) - \phi(\tau)$, these are found to be
\bes\label{PPcontractions}
\bea
I^{(4)}_{ijk}(t)I^{(5)}_{jk}(\tau) &=& \frac{16}{5}\dm\frac{\eta^2}{m^2}x^{17/2} \times \nn \\
&& \Big\{202\cos (2\varphi)\,\hbv(t) - 203 \sin (2\varphi) \,\hn(t)\Big\}_i, \nn \\
&& \\
I^{(3)}_{jk}(t)I^{(6)}_{ijk}(\tau) &=& \frac{2}{5}\dm\frac{\eta^2}{m}x^{17/2} \times \nn \\
&& \Big\{\big[-\cos (\varphi) +  3645\cos (3\varphi) \big]\hbv(t) + \nn \\
&& \big[\sin (\varphi)  + 3645\sin (3\varphi) \big] \hn(t)\Big\}_i,  
\eea
\bea
\epsilon_{ijk}I^{(3)}_{jl}(t)J^{(5)}_{kl}(\tau) &=& -2\dm\frac{\eta^2}{m^2}x^{17/2} \times \nn \\ 
&& \Big\{\cos (\varphi)\,\hbv(t) - \sin (\varphi) \, \hn(t)\Big\}_i, \nn \\ && \\ 
\epsilon_{ijk}J^{(3)}_{kl}(t)I^{(5)}_{jl}(\tau) &=& -8\dm\frac{\eta^2}{m^2}x^{17/2} \times \nn \\
&& \Big\{\cos (2\varphi)\,\hbv(t) + \sin (2\varphi) \, \hn(t)\Big\}_i. \nn \\
\eea
\ees
It is well known~\cite{BlanchetSchafer} that even though the tail integrals extend throughout the entire history of the binary, it is sufficient here to use the instantaneous Newtonian dynamics of the binary neglecting spin effects and radiation-reaction (adiabatic approximation) in order to evaluate the tails. Thus we may substitute $\delta \phi = \bar{\omega} t - \bar{\omega} \tau$ in Eqs.~\eqref{PPcontractions}, with the orbital frequency $\omega$ assumed constant. To evaluate the $\tau$ integrals, one only needs the following formula
\be\label{keyint}
\int_0^\infty \! \ln \left(\frac{u}{2B}\right) \, e^{i n \bar{\omega} u} \, du = \frac{-1}{n\bar{\omega}}\left\{\frac{\pi}{2} + i\Big[\ln(2Bn\bar{\omega}) + 
\gamma_{\rm E} \Big]\right\},
\ee
where $\gamma_{\rm E}$ is the Euler-Mascheroni constant. We provide a derivation of this essential expression for the unfamiliar reader in Appendix \ref{IntTail}. The scale $B$ in the logarithm kernel for each tail integral appearing in Eq.~\eqref{PPtail} is equal to $be^{-11/12}$, $be^{-97/60}$, $be^{-7/6}$ and $be^{-11/12}$ respectively. Performing the integrations and collecting terms yields the result
\be\label{BQWtail1}
\left(\frac{d\bm{P}}{dt}\right)_\mathrm{NS \,\, tail} \!\! = \dot{P}_N\dm x^{3/2}\Bigg\{\frac{309\pi}{58} \hbv + \ln\left(\frac{\bar{\omega}^2}{\hat{\omega}_\mathrm{NS}^2}\right)\hn\Bigg\},
\ee
where
\be\label{BQWtail2}
\hat{\omega}_\mathrm{NS} = \frac{1}{b}\exp\bigg\{\frac{5921}{1740} + \frac{48}{29}\ln 2 - \frac{405}{116}\ln 3 - \gamma_{\rm E}\bigg\}.
\ee
Equations~\eqref{BQWtail1} and~\eqref{BQWtail2} match the results of BQW. We will discuss the terms proportional to the logarithm of the orbital frequency in more details shortly. We move on to the computation of the tail terms linear in the spins. We first provide the index contractions of the relevant terms, which are
\bes
\bea
\epsilon_{ijk}I^{(3)}_{jl}(t)J^{(5)}_{kl}(\tau) &=& 3\frac{\eta^2}{m^4}x^9 \Big\{\cos (\varphi)\,(\hn\times\D^{\rm c})_i + \nn \\ &&   \sin (\varphi)\, (\hbv\times\D^{\rm c})_i 
 + \big[ \cos (\varphi)\, (\hbv\cdot\D^{\rm c}) \nn \\ &&  - \sin (\varphi)\, (\hn \cdot \D^{\rm c})\big](\hLN)_i \Big\}, \\
\epsilon_{ijk}J^{(3)}_{kl}(t)I^{(5)}_{jl}(\tau) &=& 12\frac{\eta^2}{m^4}x^9 \Big\{\cos (2\varphi)\,(\hn\times\D^{\rm c})_i \nn \\ && - \sin (2\varphi)\,(\hbv\times\D^{\rm c})_i  + \nn \\ && \big[\cos (2\varphi)\, (\hbv\cdot\D^{\rm c}) \nn \\ &&  + \sin (2\varphi)(\hn \cdot \D^{\rm c})\big](\hLN)_i \Big\}.
\eea
\ees
Again using the adiabatic approximation, which here also assumes that the spins are kept constant, the errors made being of 1PN relative order from the spin precession equations, the tails integrals can be computed immediately. The results are 
\bea
\left(\frac{d\bm{P}}{dt}\right)_\mathrm{\!SO \,\, tail}\!\!\! &=& \dot{P}_\mathrm{N}\frac{7x^2}{29m^2}\Bigg\{3\pi \Big[(\hbv\cdot\D^{\rm c})\hLN + (\hn\times\D^{\rm c})\Big] \nn \\ && \!\!\! + \ln\left(\frac{\bar{\omega}^2}{\hat{\omega}_\mathrm{SO}^2}\right)\Big[(\hn\cdot\D^{\rm c})\hLN -(\hbv\times\D^{\rm c})\Big]\Bigg\}, \nn \\ 
\eea 
where 
\be
\hat{\omega}_\mathrm{SO} = \frac{1}{b}\exp\bigg\{\frac{2}{3} - 3\ln 2 - \gamma_{\rm E}\bigg\}.
\ee
The leading spin-orbit contribution to the linear momentum flux for circular orbits is [see Eqs.~(\ref{PSlambda}), (\ref{PSLN})]
\be
\left(\frac{d\bm{P}}{dt}\right)_\mathrm{SO} = \dot{P}_\mathrm{N} \frac{7x^{1/2}}{29m^2}\Big[(\hbv\cdot\D^{\rm c})\hLN + (\hn\times\D^{\rm c})\Big],
\ee
which allows us to rewrite the spin-orbit tail as
\bea
\left(\frac{d\bm{P}}{dt}\right)_\mathrm{SO \,\, tail} &=&  3\pi x^{3/2} \left(\frac{d\bm{P}}{dt}\right)_\mathrm{SO}  \nn \\ && - x^{3/2}\ln\left(\frac{\bar{\omega}^2}{\hat{\omega}_\mathrm{SO}^2}\right) \partial_\phi \left(\frac{d\bm{P}}{dt}\right)_\mathrm{SO}, \nn \\
\eea
where the derivative $\partial_\phi$ refers to parametrization of the vectors $\hn$ and $\hbv$ in terms of the orbital phase displayed in Eqs.~\eqref{nvparam}. It becomes clear that the tail terms logarithmic in frequency can be absorbed in the leading-order spin-orbit flux by reparametrizing $\hn$ and $\hbv$ with a different phase variable $\psi_\mathrm{SO}$ defined as
\bea
\psi_\mathrm{SO}&=& \phi - 2\frac{m\bar{\omega}}{c^3}\ln\left(\frac{\bar{\omega}}{\hat{\omega}_\mathrm{SO}}\right), \label{psiSO}
\eea
where we have displayed explicitly the PN scaling of the phase modulation. The crucial point is to realize that the phase modulation induced by the tail terms is a 4PN relative correction to the orbital phase, as one can verify by taking a time derivative of Eq.~\eqref{psiSO}. Indeed one finds
\be
\dot{\psi}_\mathrm{SO} = \omega - \frac{2m}{c^3}\left[\ln\left(\frac{\bar{\omega}}{\hat{\omega}_\mathrm{SO}}\right) + 1\right]\dot{\bar{\omega}}.
\ee
Since $\dot{\bar{\omega}} \sim c^{-5}$, the second term above scales as $c^{-8}$, which shows explicitly that it is a 4PN relative correction. Since we work only at 2PN order in this paper, we ignore this phase modulation henceforth. A similar argument is made in BQW regarding the terms that are logarithmic in frequency in the non-spinning tails, i.e. they can be absorbed into a 4PN phase modulation in the leading order non-spinning linear momentum flux. This completes our discussion of the tail contributions to the linear momentum flux at 2PN order. 

\subsection{Estimate of kick velocity}

We now estimate the kick velocity of spinning black hole binaries 
moving along quasi-circular orbits using the 2PN linear momentum flux computed in the 
previous section. Since we work at 2PN accuracy, we may consider the
orbital frequency as constant when integrating the momentum flux in all terms
except the Newtonian momentum flux. When integrating the Newtonian momentum flux,
one must use Eq.~\eqref{QCorbit2} for the orbital frequency.  
Next, since the precession equations are of 
the form ${\cal O}(2) + {\cal O}(3)$, we cannot ignore the time dependence of
the spins in the ${\cal O}(1)$ spin-orbit linear momentum flux, as that time
dependence generates extra terms of order ${\cal O}(3)$ and ${\cal O}(4)$. 
On the other hand, the spins may be considered as constant in the ${\cal
  O}(3)$ and ${\cal O}(4)$ terms of the linear momentum flux, since the
precession equations generates extra terms scaling at least as ${\cal
  O}(5)$. Similarly we cannot ignore the time dependence of $\hLN$ due to precession
in the ${\cal O}(1)$ momentum flux, but it can be dropped in the ${\cal O}(3)$ and ${\cal O}(4)$
pieces.

To required accuracy, the (indefinite) integrals involving $\hn$ and $\hbv$ that
are needed for computing the kick velocity are the following  

\bes \bea
\label{rules1}
\int \hbv^i \, dt &=& \frac{1}{\bar{\omega}}\left\{1 - \frac{1}{12}\left[(\hn\cdot\Sp_0^{\rm c})^2 - (\hbv \cdot\Sp_0^{\rm c})^2 \right]\frac{x^2}{m^4}\right\}\hn^i \nn \\
&& - \frac{1}{3\bar{\omega}}(\hn\cdot\Sp_0^{\rm c})(\hbv\cdot\Sp_0^{\rm c}) \frac{(m\bar{\omega})^{4/3}}{m^4}\, \hbv^i + \mathcal{O}(5),\label{intlambda} \nn \\
\eea
\bea
\int \hn^{ijk} \, dt &=& -\frac{1}{3\bar{\omega}}\Big[\hn^{ij}\hbv^k + \hn^{ik}\hbv^j + \hn^{jk}\hbv^i\Big] \nn \\ && - \frac{2}{3\bar{\omega}}\hbv^{ijk} + \mathcal{O}(3),
\eea
\bea
\int \hbv^{ijk} \, dt &=& \frac{1}{3\bar{\omega}}\Big[\hbv^{ij}\hn^k + \hbv^{ik}\hn^j + \hbv^{jk}\hn^i\Big] \nn \\ &&  + \frac{2}{3\bar{\omega}}\hn^{ijk} + \mathcal{O}(3),
\eea
\bea
\int \hn^{ij}\hbv^k \, dt &=& \frac{1}{3\bar{\omega}} \hn^{ijk} - \frac{1}{3\bar{\omega}}\Big[\hbv^{jk}\hn^i + \hbv^{ik}\hn^j\Big]   \nn \\ && + \frac{2}{3\bar{\omega}}\hbv^{ij}\hn^k +  \mathcal{O}(3),
\eea
\bea
\int \hbv^{ij}\hn^k \, dt &=& -\frac{1}{3\bar{\omega}} \hbv^{ijk} + \frac{1}{3\bar{\omega}}\Big[\hn^{jk}\hbv^i
+ \hn^{ik}\hbv^j\Big] \nn \\ && -
\frac{2}{3\bar{\omega}}\hn^{ij}\hbv^k + \mathcal{O}(3).
\label{rules6}  \eea \ees 
Equation~\eqref{intlambda}, required for integrating the Newtonian flux, is obtained by integrating by parts using the exact expression $\hbv = \omega^{-1}(t) \dot{\hn}$ and using Eq.~\eqref{QCorbit2} for $\omega(t)$. 

The integrals involving the spin-orbit flux at 0.5PN order are of the form [see Eqs.~(\ref{PSlambda}) and (\ref{PSLN})]
\bea
\int \hbv^i \D_{\rm c}^j \, \hLN^k dt &=& \frac{1}{\bar{\omega}}\,\hbn^i \D_{\rm c}^j \, \hLN^k - 
\int \frac{1}{\bar{\omega}}\,\hbv^i\, \dot{\D}_{\rm c}^j \, \hLN^k dt \nn \\ && - \int \frac{1}{\bar{\omega}}\,\hbv^i\, \D_{\rm c}^j \, \dhLN^k dt\,+ \mathcal{O}(4).
\label{int}
\eea 
One can then substitute the evolution equation for $\hLN$, 
\be
\dhLN^i = -\frac{x^{1/2}}{\eta m^2}\,\dot{\bm{S}}_{\rm c}^i\,,
\ee
and Eqs.~\eqref{Sconstevol} and \eqref{deltaconstevol} for $\bm{S}^{\rm c}$ and 
$\D^{\rm c}$, respectively, in the integrals (\ref{int}) and perform them explicitly by
keeping the spins constant and using Eqs.~(\ref{rules1})--(\ref{rules6}).  The kick velocity $\bm{V}$ is obtained
from the following expression, \be \bm{V} =
-\frac{1}{m}\int_{-\infty}^t \dot{P} \, dt.  \ee Defining the overall
multiplicative factor \be V_\mathrm{N} = \frac{464 \eta^2
  x^4}{105} \ee so that $\bm{V} =
V_\mathrm{N}\hat{\bm{V}}$, we can split the kick velocity into the
following non-spinning and spin contributions (including  the tail terms)
\bea
\hat{\bm{V}}_\mathrm{NS} &=& \left(\dm\right)\Bigg\{1 -
\bigg(\frac{452}{87} + \frac{1139}{522}\eta\bigg)x +
\frac{309\, \pi}{58} x^{3/2} + \nn \\ && \bigg(-\frac{71345}{22968} +
\frac{36761}{2088}\eta +
\frac{147101}{68904}\eta^2\bigg)x^2\Bigg\}\hn, \nn \\ \eea
where we also included the non-spinning tail term from Ref.~\cite{BQW}, and 
\begin{widetext}
\bes
\label{FullKick} 
\bea
\hat{\bm{V}}_\mathrm{S}^{\hn} &=& \frac{x^{1/2}}{m^2}\Bigg\{-\frac{7}{29}\big[1 + 3\pi x^{3/2}\big](\hLN\cdot\D^{\rm c}) + \Bigg[-\frac{470}{87}\dm (\hLN\cdot\Sp^{\rm c}) + \left(-\frac{67}{58} + \frac{206}{29}\eta\right)(\hLN\cdot\D^{\rm c})\Bigg]x \nn \\
&& + \Bigg[ 10\dm (\hLN\cdot\Sp^{\rm c})^2 + \left(11 - 40\eta\right)(\hLN\cdot\Sp^{\rm c})(\hLN\cdot\D^{\rm c}) +\dm\left(\frac{175}{58} -10\eta\right)(\hLN\cdot\D^{\rm c})^2 \nn \\
&&  - {\frac{224}{29}}\dm (\hbv\cdot\Sp^{\rm c})^2 + \left(-{\frac{3301}{348}} + {\frac{896}{29}}\eta\right)(\hbv\cdot\Sp^{\rm c})(\hbv\cdot\D^{\rm c})  + \dm\left(-{\frac{1019}{348}} + {\frac{224}{29}}\eta\right)(\hbv\cdot\D^{\rm c})^2 \nn \\
&& - {\frac{66}{29}}\dm (\hn\cdot\Sp^{\rm c})^2 + \left({\frac{157}{87}} + {\frac{264}{29}}\eta\right)(\hn\cdot\Sp^{\rm c})(\hn\cdot\D^{\rm c}) + \dm\left({\frac{134}{87}} + {\frac{66}{29}}\eta\right)(\hn\cdot\D^{\rm c})^2\Bigg]\frac{x^{3/2}}{m^2}\Bigg\}\,.
\label{nkick}
\eea
\bea
\hat{\bm{V}}_\mathrm{S}^{\hbv} &=& \frac{x^2}{m^4}\Bigg\{{\frac{823}{29}}\dm (\hbv\cdot\Sp^{\rm c})(\hn\cdot\Sp^{\rm c}) + \left({\frac{5039}{348}} - {\frac{1646}{29}}\eta\right)(\hbv\cdot\D^{\rm c})(\hn\cdot\Sp^{\rm c}) \nn \\
&& + \left({\frac{3433}{174}} - {\frac{1646}{29}}\eta\right)(\hbv\cdot\Sp^{\rm c})(\hn\cdot\D^{\rm c}) + \dm\left({\frac{1759}{174}} - {\frac{823}{29}}\eta\right)(\hbv\cdot\D^{\rm c})(\hn\cdot\D^{\rm c})\Bigg\},
\eea
\bea
\hat{\bm{V}}_\mathrm{S}^{\hLN} &=& \frac{x^{1/2}}{m^2}\Bigg\{\frac{14}{29}\big[1+3\pi(m\omega)\big](\hn\cdot\D^{\rm c}) + \Bigg[\frac{193}{116}\dm (\hn\cdot\Sp^{\rm c}) + \left(\frac{109}{116} - \frac{127}{58}\eta\right)(\hn\cdot\D^{\rm c})\Bigg]x \nn \\
&& + \Bigg[-\frac{45}{29}\dm (\hLN\cdot\Sp^{\rm c})(\hn\cdot\Sp^{\rm c}) + \left(\frac{901}{116} + \frac{90}{29}\eta\right)(\hLN\cdot\D^{\rm c})(\hn\cdot\Sp^{\rm c}) \nn \\
&& + \left(-\frac{253}{58} + \frac{90}{29}\eta\right)(\hLN\cdot\Sp^{\rm c})(\hn\cdot\D^{\rm c}) + \dm\left(\frac{125}{58} + \frac{45}{29}\eta\right)(\hLN\cdot\D^{\rm c})(\hn\cdot\D^{\rm c})\Bigg]\frac{x^{3/2}}{m^2}\Bigg\}\,,\label{LNkick}
\eea
\ees
\end{widetext}

\subsection{Special binary configurations}

We investigate here special mass and spin configurations for which the 
recoil velocity has been computed in numerical simulations.  

The recoil velocity (\ref{FullKick}) we calculated within the PN formalism 
refers {\it only} to that portion of the total recoil accumulated 
during the inspiral phase. As shown in numerical simulations
~\cite{Baker2,Pollney,Koppitz,Herrmann,Gonzalez,Campanelli,
Bruegmann,Baker,Lousto1,Lousto2,Lousto3,recoilFAU}, and 
predicted analytically in Ref.~\cite{DG} within the effective-one-body 
model~\cite{EOB1,EOB2}, the majority of the recoil velocity is produced 
during the plunge, merger and ring-down phases. Quite interestingly, 
depending on the black holes' mass and spin, 
the integrated recoil velocity can reach a {\it peak} value (around merger) 
before decreasing to a {\it final}, smaller velocity
asymptotically. The difference between the final kick and the kick 
at the peak is generally denoted as {\it anti-kick}. Reference~\cite{anatomy} 
showed that the amount of anti-kick depends on the way the different modes 
of the linear momentum flux combine either constructively or destructively 
during the ring-down phase. While Eq.~(\ref{FullKick}) only applies to the inspiral portion, 
if anything, pushing Eq.~(\ref{FullKick}) until the merger might still 
give a rough estimate of the recoil velocity at the peak, which 
is not necessarily the same as the final, total recoil. 

By contrast, if we are interested in predicting analytically and with high accuracy 
the total recoil velocity we cannot rely on the PN-expanded 
equations (\ref{FullKick}). We would need to resum 
the linear-momentum flux or the multipole moments and build non-perturbative 
expressions which capture the correct results until merger, and 
augment them by the ringdown phase. This approach is followed in the 
effective-one-body model~\cite{EOB1,EOB2,DG}.

\subsubsection{Spins collinear with orbital angular momentum}
\label{subsubcollinear}

If the spins are collinear with the orbital angular momentum, then the projections of the spins along 
$\hn$ and $\hbv$ vanish, leaving
\bes\label{KickSpinCollinear} 
\bea
\hat{\bm{V}}_\mathrm{S}^{\hbv} &=& 0, \\
\hat{\bm{V}}_\mathrm{S}^{\hLN} &=& 0,
\eea
\begin{widetext}
\bea
\hat{\bm{V}}_\mathrm{S}^{\hn} &=& \frac{x^{1/2}}{m^2}\Bigg\{-\frac{7}{29}\big[1 + 3\pi x^{3/2}\big](\hLN\cdot\D^{\rm c}) + \Bigg[-\frac{470}{87}\dm (\hLN\cdot\Sp^{\rm c}) + \left(-\frac{67}{58} + \frac{206}{29}\eta\right)(\hLN\cdot\D^{\rm c})\Bigg]x \nn \\
&& + \Bigg[ 10\dm (\hLN\cdot\Sp^{\rm c})^2 + \left(11 -
  40\eta\right)(\hLN\cdot\Sp^{\rm c})(\hLN\cdot\D^{\rm c})
+\dm\left(\frac{175}{58} -10\eta\right)(\hLN\cdot\D^{\rm
  c})^2\Bigg]\frac{x^{3/2}}{m^2}\Bigg\}\,.  
\eea \end{widetext} \ees 
For such configurations the total kick velocity lies entirely along
$\hn$. Let us specialize Eq.~\eqref{KickSpinCollinear} to an equal
mass binary $\delta m = 0, \, \eta = 1/4$, for which the
individual spins are equal in magnitude but opposite in direction,
such that $\Sp^{\rm c} = 0$ and $|\D^{\rm c}|/m^2 = \chi = $
dimensionless spin of each individual hole. Since $\delta m = 0$ the
non-spinning and spin-spin kick contributions vanish, and the total kick reduces to 
\be
\bm{V}_\mathrm{kick} = \pm \chi\, x^{9/2}\, \Bigg\{\frac{1}{15} - \frac{6}{35}x + 
\frac{\pi}{5}x^{3/2}\Bigg\}\hn, 
\label{Vkick1}
\ee
where $\pm$ denotes whether $\D^{\rm c}$ is aligned or antialigned with $\hLN$.  
The recoil velocity for this binary configuration has been computed in several numerical 
simulations~\cite{Baker2,Pollney,Koppitz,Baker}; since it has negligible anti-kick (e.g., see Fig. 2 
in Ref.~\cite{Pollney}), the recoil velocity at the peak is close to the final, total recoil.  
The latter was estimated to be $\sim \pm \chi \,450$ km/sec. As explained above, Eq.~(\ref{Vkick1}) 
can predict only the recoil velocity around the peak. Since the choice of orbital frequency at which to compute Eq.~(\ref{Vkick1}) is rather arbitrary, we choose the range 
$m \bar{\omega} = x^{3/2} \sim 0.15\mbox{--} 0.25$, spanning the orbital frequencies when 
in the effective-one-body model the ring-down phase is joined to the inspiral phase. 
The location of the latter depends on the black holes' mass and spin. 
From Eq.~(\ref{Vkick1}) we obtain $|\bm{V}_\mathrm{kick}| \sim \chi\, 114 \mbox{--} 730$ km/sec, 
which brackets the numerical result, but has large deviations from it.

In Ref.~\cite{Pollney} numerical simulations were also carried out
relaxing the condition $\Sp^{\rm c} = 0$. In this case the magnitude
for the kick velocity is given by
\bea
|\bm{V}_\mathrm{kick}| &=& \frac{1}{2}|\chi_1 -
\chi_2|x^{9/2}\Bigg\{\frac{1}{15} - \frac{6}{35}x \nn \\ && + 
\bigg[\frac{\pi}{5} - \frac{29}{420}(\chi_1 +
\chi_2)\bigg]x^{3/2}\Bigg\}\,.  \label{Vkick1prime}
\eea
Notice here that there are
non-zero spin-spin contributions to the kick. In Ref.~\cite{Pollney},
it is demonstrated that while the measured recoil velocities are not
very well approximated by the original Kidder kick
formula~\cite{Kidder}, the addition of terms quadratic in spin to the
Kidder fitting formula provides very good matching to numerical
data. The fitting function of Ref.~\cite{Pollney} has the following
form \be\label{PollneyFit1} |\bm{V}_\mathrm{kick}| =
|\chi_2|f\left(\frac{\chi_1}{\chi_2}\right), \ee where
\be\label{PollneyFit2} f(y) = 109.3 - 132.5\,y + 23.1 \, y^2 \,\,
\mathrm{km/s}.  \ee 
Given a fixed $\chi_2$, the authors of Ref.~\cite{Pollney} show that
Eqs.~\eqref{PollneyFit1}--\eqref{PollneyFit2} help to reduce the fit
residuals from 20 km/s (Kidder formula) to 5 km/s. However it should
be pointed out here that the functional form of
Eqs.~\eqref{PollneyFit1}--\eqref{PollneyFit2} is not invariant under
interchange of particle labels\footnote{Note however that
  Eq.~\eqref{Vkick1prime} clearly is invariant under a relabeling of
  the black holes.}. As was highlighted in Ref.~\cite{SpinExpansion},
this fundamental symmetry provides a guiding principle for building a
viable kick velocity fitting formula over the entire binary parameter
space. The fit provided in Eq.~\eqref{PollneyFit2} is derived from a
series of simulations where $\chi_2$ is kept constant (equal to
-0.584) and $\chi_1$ is varied. Thus, there must remain some hidden
dependence on $\chi_2$ in the numerical coefficients in
Eq.~\eqref{PollneyFit2} so that particle-label's symmetry is
satisfied. Indeed our expression~\eqref{Vkick1prime} suggests the
following alternative fitting formula \bea\label{ImprovedFit}
|\bm{V}_{\mathrm{kick}}| &=& |\chi_1 - \chi_2|\big[d_1 + d_2(\chi_1 + \chi_2)\big] \nn \\
&=& |\chi_2|\big[d_1(1-y) + \chi_2 d_2 (1-y^2)\big], \eea where $y =
\chi_1/\chi_2$ (with $|y|<1$), and where $d_1$ and $d_2$ are constants
determined from the data of Ref.~\cite{Pollney}, the results being
\bes\label{dFitVals} \bea
d_1 &=& 226.9 \, \mathrm{km/s}, \\
d_2 &=& 67.8 \, \mathrm{km/s}.  \eea \ees Note that this modified fit
does not change the maximum kick value of $\sim 450$ km/s, obtained
when $\chi_1 = -\chi_2 = \pm 1$. Nevertheless 
Eqs.~\eqref{ImprovedFit}--\eqref{dFitVals} may provide better
results than Eqs.~\eqref{PollneyFit1}--\eqref{PollneyFit2} when both
$\chi_1$ and $\chi_2$ are varied.

\subsubsection{Spins perpendicular with orbital angular momentum}
\label{sec:Superkick}

Let us now investigate the so-called superkick configuration studied in
several numerical simulations~\cite{Campanelli,Gonzalez,recoilFAU,Baker,
Lousto1,Lousto2,Lousto3}. We specialize Eq.~\eqref{FullKick} to
the case of an equal mass binary $\delta m = 0, \, \eta = 1/4$, 
for which the individual spins are equal in magnitude but opposite in
direction, i.e., $\Sp^{\rm c} = 0$ and $|\D^{\rm c}|/m^2 = \chi =
$ dimensionless spin of each individual hole. The spins lie 
initially on the orbital plane. For this 
particular spin configuration, the precession equations
\eqref{Sconstevol} and \eqref{deltaconstevol} ensure that the total 
spin remains zero and that $\D^{\rm c}$ remains in the orbital plane for all
time, precluding precession of the orbital plane. 
In this case the non-spinning kick vanishes, as
well as the spin contributions along $\hn$ and $\hbv$, leaving the
contribution along $\hLN$ as the lone non-zero term. The total kick
velocity is thus entirely out of the orbital plane and is equal to
\be
\label{NOsuperkick} 
\bm{V}_\mathrm{kick} = -\chi\,\cos\varphi\,x^{9/2}\,\Bigg\{\frac{2}{15} +
\frac{13}{120}x + \frac{2\pi}{5}x^{3/2}\Bigg\} \hLN,
\ee 
where $\varphi$ is the angle between $\D^{\rm c}$ and $\hn$. For this 
binary configuration, the anti-kick is absent (e.g., see Fig. 17 
in Ref.~\cite{anatomy}). Thus the recoil velocity at the peak (around merger) 
is close to the final, total recoil. As done 
in Sec.~\ref{subsubcollinear}, we estimate the recoil velocity at the 
peak from Eq.~(\ref{NOsuperkick}) varying the orbital 
frequency in the range $m \bar{\omega} \sim 0.15\mbox{--} 0.25$. 
Maximizing on $\varphi$ we obtain $|\bm{V}_\mathrm{kick}| \sim \chi\, 357 \mbox{--} 2300$ km/sec.  
Thus, even the maximum value obtained for maximal spins $2300$ km/sec is 
somewhat below the value $4000\,\mathrm{km/s}$ predicted by the 
numerical simulation. [Note that due to the fast increase of the recoil velocity 
at high frequency, had we computed the recoil at $m \bar{\omega} \sim 0.3$, 
we would have obtained $4527$ km/sec.] While Eq.~(\ref{NOsuperkick}) for the
kick velocity might not be quite trustworthy at such high orbital
frequencies, it is worth to note that the higher-order spin terms computed in this 
paper increase the recoil velocity by a factor $\sim 1.5-2.7$ with respect to the leading 
order spin term computed by Kidder~\cite{Kidder}, i.e., with respect to 
the $2/15$ term in Eq.~\eqref{NOsuperkick}. Notice also that the kick predicted 
by PN theory for this superkick configuration is linear in the spins, i.e. 
all spin-spin terms vanish in that configuration. 

It is interesting to note that the quasi-circular radiation-reaction force 
in the superkick configuration could be deduced from Eqs.~(\ref{dPdtSOcoeffs}), 
(\ref{dPdtNLSOcoeffs}) using linear-momentum balance arguments. For example at 
leading PN order, Eq.~(\ref{dPdtSOcoeffs}) says that the radiation-reaction 
force is normal to the orbital plane and changes sign as the spins precess 
on the orbital plane. It reaches its maximum value when the spins are 
collinear with the instantaneous orbital velocity vector $\hbv$, and 
it is zero when the spins are perpendicular to $\hbv$. This radiation-reaction 
force causes the binary center-of-mass to oscillate with increasing amplitude up 
and down along the direction perpendicular to the initial position of 
the orbital plane. The magnitude and direction of the recoil 
velocity normal to the orbital plane is ultimately determined by where 
in the orbit the end of the inspiral (i.e., the merger) occurs. This picture 
was also suggested in Ref.~\cite{FP} (see Fig. 5 therein) although there the argument 
is constructed using the conservative dynamics. Because of that reason it is not clear to us 
how the picture of Ref.~\cite{FP} carries over to the radiation-reaction force. However
since both Ref.~\cite{FP} and ourselves arrive (qualitatively) the same result, we suspect
there must exist a relationship between the radiation-reaction force driving the recoil
and the conservative force responsible for frame dragging, although we are unaware of any
explicit formulation of that correspondance.

\subsubsection{Out-of-plane kick for generic configurations}

Here we rewrite Eq.~\eqref{LNkick} in terms of individual dimensionless 
spins $\bc_A = \Sp_A / m_A^2$ (omitting the ${\,}^{\rm c}$ superscripts here 
for sake clarity of notation below) to provide a formula that can be used 
when comparing to numerical simulations, and also to shed some light 
in the recent controversy of whether the recoil velocity 
out-of-plane scales like $\eta^2$~\cite{Lousto3} or $\eta^3$~\cite{Campanelli,Lousto3}.

If we define the angles $\Theta$ and $\Psi$ as follows 
\bes \bea
\hn \cdot \D &=& \Delta^\perp \cos\Theta ,\\
\hn \cdot \Sp &=& S^\perp \cos\Psi , \eea \ees where \bes \bea
\Delta^\perp &=& |\D - (\hLN\cdot\D)\hLN| \nn \\
&=& \frac{m^2}{(q+1)}|\bc^\perp_2 - q\bc^\perp_1| , \eea \bea
S^\perp &=& |\Sp - (\hLN\cdot\Sp)\hLN| \nn \\
&=& \frac{m^2}{(q+1)^2}|\bc_2^\perp + q^2 \bc^\perp_1|, \eea \ees 
where $\bc_A^\perp = \bc_A - (\hLN\cdot\bc_A)\hLN$, then
the component of the kick along the orbital angular momentum axis can
be rewritten as
\bea 
\hLN \cdot \bm{V}_\mathrm{kick} &=&
K\cos(\Theta-\pi) \frac{\eta^2}{(q+1)} |\bc^\perp_2 - q\bc^\perp_1| \nn \\ &&  +
K^\prime \cos (\Psi - \pi) \frac{\eta^2}{(q+1)^3}|\bc_2^\perp + q^2
\bc^\perp_1|, \nn \\ 
\eea 
where 
\bea
K &=& x^{9/2}\,\Bigg\{\frac{32}{15}\,\Big[1 + 3\pi x^{3/2}\Big] + 4\left(\frac{109}{105} - \frac{254}{105}\eta\right)x \nn \\
&&  + 8\left[\left(-\frac{253}{105} + \frac{12}{7}\eta\right)(\chi_2^\parallel +q^2\chi_1^\parallel) \right.\nn \\ && \left. + (q-1)\left(\frac{25}{21} + \frac{6}{7}\eta\right)(\chi_2^\parallel - q\chi_1^\parallel)\right]\frac{x^{3/2}}{(q+1)^2}\Bigg\}\,, \nn \\ && \\ 
K^\prime &=&x^{11/2}\Bigg\{(q-1)\left[\frac{772}{105} -
  \frac{48}{7}\frac{(\chi_2^\parallel
    +q^2\chi_1^\parallel)}{(q+1)^2}x^{1/2}\right] \nn \\ && +
\left(\frac{3604}{105} + \frac{96}{7}\eta\right)(\chi_2^\parallel -
q\chi_1^\parallel)x^{1/2}\Bigg\}.  
\eea 
The quantities $\chi_A^\parallel$ are equal to $\bc_A \cdot\hLN$. Thus, the out-of-plane 
recoil velocity has a non trivial dependence on $\eta$ when higher order PN corrections 
are included. The dominant contribution scales as $\eta^2$, but there are additional 
non-negligible contributions scaling as $\eta^3$.  

\section{Energy and angular momentum fluxes}
\label{sec:EJfluxes}

To provide further checks on our methodology, we provide here a (complementary) computation of the energy and angular momentum 
fluxes at 2PN order for spinning binary black holes, including all spin contributions. The flux formulas, taken from Thorne \cite{Thorne}, are
\bes\label{fluxes}
\bea
\frac{dE}{dt} &=& \frac{1}{5}I^{(3)}_{ij}I^{(3)}_{ij} + \frac{1}{c^2}\left[\frac{1}{189}I^{(4)}_{ijk}I^{(4)}_{ijk} + \frac{16}{45}J^{(3)}_{ij}J^{(3)}_{ij}\right] \nn \\ 
&& + \frac{1}{c^4}\left[\frac{1}{9072}I^{(5)}_{ijkl}I^{(5)}_{ijkl} + \frac{1}{84}J^{(4)}_{ijk}J^{(4)}_{ijk}\right] \nn \\
&& + \frac{1}{c^3}\left(\frac{dE}{dt}\right)_{\rm tail} , \label{Eflux}\\
\frac{dJ_i}{dt} &=& \epsilon_{ijk}\left\{\frac{2}{5}I^{(2)}_{jl}I^{(3)}_{kl} + \frac{1}{c^2}\left[\frac{1}{63}I^{(3)}_{jlm}I^{(4)}_{klm} + \frac{32}{45}J^{(2)}_{jl}J^{(3)}_{kl}\right] \right. \nn \\
&& + \frac{1}{c^4}\left[\frac{1}{2268}I^{(4)}_{jlmn}I^{(5)}_{klmn} + \frac{1}{28}J^{(3)}_{jlm}J^{(4)}_{klm}\right] \nn \\ && \left. + \frac{1}{c^3}\left(\frac{dJ_i}{dt}\right)_{\rm tail}\right\}, 
\eea
\ees
where the tail terms are\footnote{In Ref.~\cite{Rieth}, expression (83) for the angular momentum tail integral is missing the ``I2I5'' term. This is only a typo, as their computations take that term into account. Equation (3.10) in Ref.~\cite{GI} is quoting Eq.~(83) of Ref.~\cite{Rieth} and carries the same typo.}
\bea
\left(\frac{dE}{dt}\right)_{\rm tail} \!\! &=& \frac{4m}{5}I^{(3)}_{ij}(t)\int_{-\infty}^t d\tau I^{(5)}_{ij}(\tau)\left[\ln\left(\frac{t - \tau}{2b\, e^{-\frac{11}{12}}}\right)\right], \label{dEdttail} \nn \\ && \\ 
\left(\frac{dJ_i}{dt}\right)_{\rm tail}\!\!  &=& \frac{4m}{5}\epsilon_{ijk}\int_{-\infty}^t d\tau\left[\ln\left(\frac{t-\tau}{2b\, e^{-\frac{11}{12}}}\right)\right]\times \nn \\ 
&& \bigg\{I^{(2)}_{jl}(t)I^{(5)}_{kl}(\tau) + I^{(3)}_{kl}(t)I^{(4)}_{jl}(\tau)\bigg\}.
\eea
We note here that throughout this section $\Sp$ and $\D$ can be freely interchanged with $\Sp^{\rm c}$ and $\D^{\rm c}$, as the differences generated by this substitution appear only at (relative) 2.5PN order in the energy and angular momentum fluxes. 

\subsection{Main results}

The non-spinning contributions up to 2PN order and spin-orbit contributions at 1.5PN order to the energy and angular momentum fluxes are all well-known. 
Our goal here is to compute the 2PN terms quadratic in sp
ins in the energy and angular momentum fluxes, so only a handful 
of terms from Eqs.~\eqref{fluxes} contribute. 
%Following the notation introduced in Eq.~\eqref{contributors}, the contributing terms for the energy flux are
%\be
%\frac{1}{c^4}\left(\frac{dE}{dt}\right)_\mathrm{SS} = \frac{2}{5}I^{(3)}_{ij}[c^{-4};\mathrm{SS}]I^{(3)}_{ij}[c^{-0};\mathrm{NS}] + \frac{16}{45\, c^2}J^{(3)}_{ij}[c^{-1};
%\mathrm{S}]J^{(3)}_{ij}[c^{-1};\mathrm{S}].
%\ee
%For the angular momentum flux, the contributing terms are
%\bea
%\frac{1}{c^4}\left(\frac{dJ_i}{dt}\right)_\mathrm{SS} = \frac{2}{5}\epsilon_{ijk}I^{(2)}_{jl}[c^{-4};\mathrm{SS}]I^{(3)}_{kl}[c^{-0};\mathrm{NS}] + 
%\frac{2}{5}\epsilon_{ijk}I^{(2)}_{jl}[c^{-0};\mathrm{NS}]I^{(3)}_{kl}[c^{-4};\mathrm{SS}] + \frac{32}{45\,c^2}\epsilon_{ijk}J^{(2)}_{jl}[c^{-1};\mathrm{S}]J^{(3)}_{kl}[c^{-1};\mathrm{S}]. 
%\nonumber \\
%\eea
Since the spin-spin acceleration depends solely on the spin combination $\Sp_0$, it turns to be much more natural to write the fluxes in terms of 
$\Sp_0$ and $\D$ instead of $\Sp$ and $\D$.~\footnote{The reason why it is not necessarily advantageous to switch from $\Sp$ to $\Sp_0$ in the case of the linear momentum flux is that the spin-spin orbital acceleration is just one contribution among a lot of other contributions that do not simplify when written in terms of $\Sp_0$. However in the case of the energy and angular momentum fluxes the spin-spin orbital acceleration is the dominant contribution (in number of terms). We find that the spin-spin terms in the energy and angular momentum fluxes simplify drastically when expressed in terms of $\Sp_0$ rather than $\Sp$.}. The results for generic orbits are
\begin{widetext}
\bea\label{dEdtSSgeneric}
\left(\frac{dE}{dt}\right)_\mathrm{SS} &=& \frac{2\eta^2 m^2}{5r^6}\Bigg\{4(12v^2 - 13\rd^2)\Sp_0^2    - 8(21v^2 -34\rd^2)(\hn\cdot\Sp_0)^2 + 24 (\bv\cdot\Sp_0)^2 - 116\rd (\hn\cdot\Sp_0)(\bv\cdot\Sp_0) \nn \\
&& + (v^2 + 3\rd^2)(\D)^2 + 3\rd^2(\hn\cdot\D)^2 + \frac{1}{3}(\bv\cdot\D)^2 - 2\rd(\hn\cdot\D)(\bv\cdot\D)   \Bigg\}, \label{dEdtSSanswer}
\eea
\bea\label{dJdtSSgeneric}
\left(\frac{dJ_i}{dt}\right)_\mathrm{SS} &=& \frac{2\eta^2m}{5r^4}\Bigg\{+ \bigg[\mr \D^2 - 90\rd (\hn\cdot\Sp_0)(\bv\cdot\Sp_0) + 6 (\bv\cdot\Sp_0)^2  - 30\bigg(2v^2 - 7\rd^2 + 3\mr\bigg)(\hn\cdot\Sp_0)^2 \nn \\
&& + 6\bigg(2v^2 - 5\rd^2 + 4\mr\bigg)(\Sp_0)^2\bigg](\hn\times\bv)_i + 6\mr\big[(\bv -\rd\hn)\cdot\Sp_0\big](\hn\times\Sp_0)_i  + \mr(\bv\cdot\D)(\hn\times\D)_i \nn \\ && - \mr(\hn\cdot\D)(\bv\times\D)_i - 6\bigg[\rd (\bv\cdot\Sp_0) + \bigg(3v^2 - 5\rd^2 + 2\mr\bigg)(\hn\cdot\Sp_0) \bigg](\bv\times\Sp_0)_i  \Bigg\}.\label{dJdtSSanswer}
\eea
\end{widetext}
\subsection{Test-mass limit}

One important check of our computations is provided by the limiting case of a test mass orbiting a Kerr black hole. In Ref.~\cite{Tagoshi}, Tagoshi {\it et al.} computed a PN expansion of the energy flux produced by a test mass in a circular equatorial orbit around a Kerr black hole obtained via the Teukolsky formalism. In this section we show that our energy flux matches the expression of \cite{Tagoshi} at 2PN order. Restricting attention to circular orbits in the equatorial plane, one can solve for the orbital angular frequency $\omega$ using Eqs.~\eqref{accs} for a non-spinning test mass, the result being 
\bea
\omega &=& \frac{1}{m} \left(\mr\right)^{3/2} \left[1 - \frac{3}{2}\left(\mr\right) - \chi\left(\mr\right)^{3/2} \right. \nn \\ 
&& \left. + \frac{3}{8}\left(5+2\chi^2\right)\mrsq\right], \label{omegaPN}
\eea
where $\chi$ is the dimensionless spin of the Kerr hole, which is denoted by $q$ in Ref.~\cite{Tagoshi}. Since it is an observable encoded in the gravitational radiation observed at null infinity, the orbital frequency is a gauge invariant quantity. We can therefore use it to relate the harmonic gauge radial coordinate $r$ of PN theory to the Boyer-Lindquist radial coordinate $r_0$ of Ref.~\cite{Tagoshi}. Defining ${\rm v} = (m/r_0)^{1/2}$, Tagoshi {\it et al.} found~
\footnote{The quantity ${\rm v}$ defined in Ref.~\cite{Tagoshi} is not to be confused with the orbital velocity $v$ of PN theory.}
\be\label{omegaTag}
\omega = \frac{1}{m} {\rm v}^3\left[1 - \chi{\rm v}^3 + {\cal O}({\rm v}^6)\right].
\ee
Equating Eqs.~\eqref{omegaPN} and~\eqref{omegaTag}, one obtains, to 2PN accuracy, the following relation
\be\label{mrofv}
\left(\mr\right)^{1/2} = {\rm v}\left[1 + \frac{1}{2}{\rm v}^2 + \frac{1}{8}\left(3 - 2\chi^2\right){\rm v}^4 + {\cal O}({\rm v}^5)\right].
\ee
Substituting $\rd = 0$ and $v = \omega r$ into Eq.~\eqref{dEdtSSanswer} for the energy flux, and supplementing the resulting expression with all other contributing terms at 2PN order (see e.g. Ref.~\cite{WS}, ignoring however Eq.~(F17) for the spin-spin energy flux, which is incomplete), one can verify straightforwardly, making use of Eqs.~\eqref{omegaTag} and~\eqref{mrofv}, that the resulting energy flux at 2PN order is
\bea
\frac{dE}{dt} &=& \frac{32\eta^2}{5}{\rm v}^{10}\left[1 - \frac{1247}{336}{\rm v}^2 + \left(4\pi- \frac{73}{12}\chi\right) {\rm v}^3 \right. \nn \\ && \left. + \left(\frac{33}{16}\chi^2 - \frac{44711}{9072}\right){\rm v}^4\right], \label{dEdtTag}
\eea
which precisely matches Eq.~(3.40) of Ref.~\cite{Tagoshi} computed from black hole perturbation theory. The $4\pi$ term at 1.5PN order is the contribution from the tail integral given by Eq.~\eqref{dEdttail}. The corresponding computation of the angular momentum flux is straightforward, the additional terms contributing at 2PN order being found in Refs.~\cite{Kidder,GI}. The result is
\bea
\frac{d\bm{J}}{dt} &=& \frac{32\eta^2m}{5}{\rm v}^{7}\left[1 - \frac{1247}{336}{\rm v}^2 + \left(4\pi- \frac{61}{12}\chi\right) {\rm v}^3 \right.\nn \\ && \left.+ \left(\frac{33}{16}\chi^2 - \frac{44711}{9072}\right){\rm v}^4\right]\hLN.
\eea
By using Eq.~\eqref{omegaTag} together with Eq.~\eqref{dEdtTag}, one can rewrite the angular momentum flux very simply as
\be
\frac{d\bm{J}}{dt} = \frac{1}{\omega} \frac{dE}{dt} \hLN,
\ee
which verifies Eq.~(3.41) of Ref.~\cite{Tagoshi}.

\subsection{Evolution of the orbital frequency}

We compute here the evolution equation for the orbital frequency, derived from the usual energy balance argument and specialized to quasi-circular orbits. 
The balance argument says that the (average) orbital frequency evolves according to
\be\label{balance}
\left\langle\frac{d\omega}{dt}\right\rangle = \left\langle \frac{dE/dt}{dE_\mathrm{orb}/d\omega}\right\rangle,
\ee
where $dE/dt$ is given by Eq.~\eqref{Eflux} and where $E_\mathrm{orb}(\omega)$ is the orbital energy. The orbital energy, which is conserved by the 2PN orbital dynamics defined by Eqs.~\eqref{accs}, is given by
\be
E_\mathrm{orb} = E_\mathrm{N} + \frac{1}{c^2} E_\mathrm{1PN} + \frac{1}{c^3} E_\mathrm{SO} + \frac{1}{c^4} E_\mathrm{2PN} + \frac{1}{c^4} E_\mathrm{SS},
\ee
where
\bes\label{EorbPN}
\bea
E_\mathrm{N} &=& \eta m\left[\frac{1}{2}v^2 - \mr\right] \\
E_\mathrm{1PN} &=& \frac{\eta m}{2}\left[\frac{3}{4}(1-3\eta)v^4 + (3+\eta)\mr v^2 \right. \nn \\ && \left. + \eta \mr \rd^2 + \mrsq\right] \\
E_\mathrm{SO} &=& \frac{1}{r^3}\bm{L}_\mathrm{N}\cdot\left[\Sp + \dm \D\right] \\
E_\mathrm{2PN} &=& \frac{\eta m}{4}\left[\frac{5}{4}(1-7\eta+13\eta^2)v^6 + \eta(1-15\eta)\mr \rd^2v^2 \right. \nn \\ && \left. + \frac{1}{2}(21-23\eta - 27\eta^2)\mr v^4 - \frac{3}{2}\eta(1-3\eta)\mr \rd^4 \right. \nn \\
&& \left. + \frac{1}{2}(14-55\eta+4\eta^2)\mrsq v^2 - (2+15\eta)\mrcb \right. \nn \\ && \left. + \frac{1}{2}(4+69\eta+12\eta^2)\mrsq \rd^2 \right]\\
E_\mathrm{SS} &=& \frac{\eta}{2r^3}\bigg[3(\hn\cdot\Sp_0)^2 - \Sp_0^2\bigg].
\eea
\ees
Reducing Eqs.~\eqref{dEdtSSgeneric} and~\eqref{EorbPN} to circular orbits following the same prescription as performed in Sec.~\ref{sec:circ} for the linear momentum flux, substituting the results into Eq.~\eqref{balance} and averaging the result over orbital motion yields
\bea
\left\langle\frac{d\omega}{dt}\right\rangle &=& \omega^2 \frac{96\eta}{5}x^{5/2}\Bigg[1 - \left(\frac{743}{336} + \frac{11}{4}\eta\right)x + 4\pi x^{3/2} \nn \\ &&   + \left(\frac{34103}{18144} + \frac{13661}{2016}\eta + \frac{59}{18}\eta^2 \right)x^2 \nn \\ &&   - \beta x^{3/2} + \sigma x^2 \Bigg],
\eea
where
\bea\label{beta}
\beta &=& \frac{47}{3m^2}\hLN\cdot{\Sp} + \frac{25}{4m^2}\dm \hLN\cdot{\D} \nn \\
&=& \frac{1}{12}\sum_A \left[113\frac{m_A^2}{m^2} + 75\eta\right]\hLN\cdot\bc_A
\eea
and where
\bea
\sigma &=& -10 \frac{\Sp^2}{m^4} - 10\dm \frac{\Sp}{m^2}\cdot\frac{\D}{m^2} + \left(-\frac{233}{96} + 10\eta\right)\frac{\D^2}{m^4} \nn \\ && + 30\left(\hLN\cdot\frac{\Sp}{m^2}\right)^2 + 30\dm \left(\hLN\cdot\frac{\Sp}{m^2}\right)\left(\hLN\cdot\frac{\D}{m^2}\right) \nn \\ && + \left(\frac{719}{96} - 30\eta\right)\left(\hLN\cdot\frac{\D}{m^2}\right)^2 \nn \\
&=& \frac{\eta}{48}\bigg[721(\hLN\cdot\bc_1)(\hLN\cdot\bc_2) - 247\bc_1\cdot\bc_2\bigg] \nn \\ && + \frac{1}{96}\sum_A\frac{m_A^2}{m^2}\left[719(\hLN\cdot\bc_A)^2 - 233\bc_A^2\right] \label{sigma}
\eea
In Eqs.~\eqref{beta} and~\eqref{sigma}, we have $\bc_A = \Sp_A/m_A^2$. The term involving the sum over $A$ in Eq.~\eqref{sigma} is generally omitted in the literature, but does indeed contribute at the same order as the $\chi_1\chi_2$ piece. It should therefore be included in templates for spinning binary black holes. Equation~\eqref{sigma} is equivalent to the sum of Eqs.~(9b),~(9c) and~(9d) of Ref.~\cite{MVG}, when the quadrupole moment of a Kerr black hole is substituted into Eq.~(9d). This completes our report on energy and angular momentum fluxes at 2PN for spinning binaries.
 
\section{Conclusions}

In this paper we computed the linear-momentum flux carried by
gravitational waves emitted from spinning binary black holes at 2PN
order for generic orbits, notably the next-to-leading order spin-orbit
terms at 1.5 PN order, spin-orbit tail terms at 2PN order, and
spin-spin terms at 2PN order. In addition, as far as we know, the 2PN
non-spinning terms for generic orbits we provide do not seem appear in
the literature. We also performed the reduction to quasi-circular
orbits and integrated the simplified flux over time to obtain the kick
velocity as function of orbital frequency. We specialized our formula
for the kick velocity of equal mass binary configurations where the
spins are equal in magnitude, opposite in direction and are either
collinear with the orbital angular momentum or lying in the orbital
plane. In particular, we found that in the so-called superkick configuration the higher-order 
  spin corrections computed in this paper can increase the recoil velocity up to a 
  factor $\sim 3$ with respect to the leading-order PN prediction.

Comparisons between PN and numerical relativity results for the 
gravitational-wave energy flux have already shown that when 
the latter is computed for quasi-circular orbits in the adiabatic 
approximation, as done in this paper for the linear-momentum 
flux, it tends to overestimate the numerical energy flux during the last 
stages of inspiral and plunge~\cite{Boyleetal08}. Similar conclusions 
can be drawn here for the linear-momentum flux.

The PN expressions for the linear-momentum flux can be used to grasp 
which asymmetry in the parameter space can produce the recoil velocity, 
or suggest phenomenological formulas describing numerical-relativity 
results~\cite{Campanelli,Herrmann,Baker,Lousto3}. 
However, not surprisingly, the fast increase during the late inspiral and plunge, 
and the arbitrariness in determining until when those formula should 
be trusted, make the PN predictions not very accurate and robust 
for predicting the recoil velocity at the peak. By contrast, the computation 
of the linear-momentum flux at higher PN orders is crucial for 
building more reliable resummed expressions aimed at capturing the 
non-perturbative effects until merger~\cite{EOB1,EOB2,DG,SB} 
and predict the total recoil velocity.

We also provided expressions valid for generic orbits, and accurate at
2PN order, for the energy and angular momentum carried by
gravitational waves emitted from spinning binary black
holes. Specializing to quasi-circular orbits we computed the
derivative of the orbital frequency through 2PN order, and found
agreement with results of Mik\'{o}czi, Vas\'{u}th and
Gergely~\cite{MVG}. We also verified that in the limit of extreme mass
ratio our expressions for the energy and angular momentum fluxes match
the results of Tagoshi {\it et al.}~\cite{Tagoshi} obtained in the
context of black hole perturbation theory.

It would certainly be quite interesting to extend this computation to
3PN order to provide more refined estimates of the recoil velocity 
accumulated during the inspiral and build more accurate resummed expressions. 
This would require the computation of several new source multipole moments. 
In addition the 3PN acceleration and 3PN spin precession
equations currently available in the literature~\cite{PortoS1S2,
PortoS1S1,SpinSchafer1,SpinSchafer2,SpinSchafer3} would need to be 
computed in harmonic gauge. 

\begin{acknowledgments}
A.B. and E.R. acknowledge support from NSF grant No. PHY-0603762, 
and A.B. also acknowledges support from the Alfred P. Sloan Foundation. 
L.K. is supported by a grant from the Sherman Fairchild Foundation, and NSF grants No. PHY-0652952, No. DMS-0553677 and No. PHY-0652929.

The results of this paper were obtained using three independent 
codes based on Mathematica and MathTensor.
\end{acknowledgments}

\appendix

\section{Source multipole moments}
\label{App:moments}

Here we list all source multipole moments required for our
computations. The expressions below are only valid in the
center-of-mass frame. The mass moments $I_{ij}$ and $I_{ijk}$, along
with the current moment $J_{ij}$ are needed at ${\cal O}(4)$ (2PN)
accuracy. The mass moment $I_{ijkl}$ and the current moment $J_{ijk}$
are required at ${\cal O}(2)$ (1PN) accuracy, and the moments
$I_{ijklm}$ and $J_{ijkl}$ are needed at ${\cal O}(0)$ (Newtonian)
accuracy. We denote with $<i \dots j>$ the symmetric trace-free part with respect to 
the indices $i$ and $j$. The explicit expressions are
\begin{widetext}
\bea
I_{ij} &=& \eta m\Bigg\{ \Bigg[1 + \frac{1}{c^2}\bigg[\left(\frac{29}{42} - \frac{29}{14}\eta\right)v^2 + \left(-\frac{5}{7} + \frac{8}{7}\eta\right)\mr\bigg]  + \frac{1}{c^4}\bigg[\left(\frac{253}{504} - \frac{1835}{504}\eta + \frac{3545}{504}\eta^2\right)v^4  \nn \\
&& + \left(\frac{2021}{756} - \frac{5947}{756}\eta - \frac{4883}{756}\eta^2\right)\mr v^2 + \left(-\frac{131}{756} + \frac{907}{756}\eta - \frac{1273}{756}\eta^2\right)\mr \rd^2  \nn \\
&& + \left(-\frac{355}{252} - \frac{953}{126}\eta + \frac{337}{252}\eta^2\right)\mrsq \bigg]\Bigg]x_{<ij>} + \frac{r^2}{c^2}\Bigg[\frac{11}{21} - \frac{11}{7}\eta + \frac{1}{c^2}\bigg[\left(\frac{41}{126} - \frac{337}{126}\eta + \frac{733}{126}\eta^2\right)v^2 \nn \\
&&+ \left(\frac{5}{63} - \frac{25}{63}\eta + \frac{25}{63}\eta^2\right)\rd^2 + \left(\frac{106}{27} - \frac{335}{189}\eta - \frac{985}{189}\eta^2\right)\mr\bigg] \Bigg] v_{<ij>} + 2\frac{r\rd}{c^2}\Bigg[-\frac{2}{7} + \frac{6}{7}\eta \nn \\
&& + \frac{1}{c^2}\bigg[\left(-\frac{11}{13} + \frac{101}{63} - \frac{209}{63}\eta^2\right)v^2 + \left(-\frac{155}{108} + \frac{4057}{756}\eta + \frac{209}{108}\eta^2\right)\mr \bigg]\Bigg]x_{<i}v_{j>}\Bigg\} \nn \\
&& + \frac{\eta}{c^3}\Bigg\{\Bigg[\frac{8}{3}x_{<i}(\bv\times\Sp)_{j>} - \frac{4}{3}v_{<i}(\bm{x}\times\Sp)_{j>} + \frac{8}{3}\dm x_{<i}(\bv\times\D)_{j>} - \frac{4}{3}\dm v_{<i}(\bm{x}\times\D)_{j>}\Bigg]\Bigg\},
\eea
\bea
I_{ijk} &=& \eta\, \delta m \Bigg\{\Bigg[-1 + \frac{1}{c^2}\bigg[\left(-\frac{5}{6} + \frac{19}{6}\eta\right)v^2 + \left(\frac{5}{6} - \frac{13}{6}\eta\right)\mr\bigg] + \frac{1}{c^4}\bigg[\left(-\frac{257}{440} + \frac{7319}{1320}\eta - \frac{5501}{440}\eta^2\right)v^4 \nn \\
&& + \left(-\frac{3853}{1320} + \frac{14257}{1320}\eta + \frac{17371}{1320}\eta^2\right)\mr v^2 + \left(\frac{247}{1320} - \frac{531}{440}\eta + \frac{1347}{440}\eta^2\right)\mr \rd^2 \nn \\
&& + \left(\frac{47}{33} + \frac{1591}{132}\eta - \frac{235}{66}\eta^2\right)\mrsq\bigg]\Bigg]x_{<ijk>} + \frac{r\rd}{c^2}\Bigg[1 - 2\eta + \frac{1}{c^2}\bigg[\left(\frac{13}{22} - \frac{107}{22}\eta + \frac{102}{11}\eta^2\right)v^2 \nn \\
&&+ \left(\frac{2461}{660} - \frac{8689}{660}\eta - \frac{1389}{220}\eta^2\right)\mr\bigg]\Bigg]x_{<ij}v_{k>}  + \frac{r^2}{c^2}\Bigg[-1 + 2\eta + \frac{1}{c^2}\bigg[\left(-\frac{61}{110} + \frac{519}{110}\eta - \frac{504}{55}\eta^2\right)v^2 \nn \\
&& + \left(\frac{1}{11} - \frac{4}{11}\eta + \frac{3}{11}\eta^2\right)\rd^2 + \left(-\frac{1949}{330} - \frac{62}{165}\eta + \frac{483}{55}\eta^2\right)\mr\bigg]\Bigg]x_{<i}v_{jk>} + \frac{r^3\rd}{c^4}\left(- \frac{13}{55} + \frac{52}{55}\eta - \frac{39}{55}\eta^2\right)v_{<ijk>}\Bigg\} \nn \\
&& + \frac{3\eta}{2} \frac{1}{c^3}\Bigg\{-3\dm x_{<ij}(\bv\times\Sp)_{k>}  + \big(-3 + 11\eta\big)x_{<ij}(\bv\times\D)_{k>} + 2\dm x_{<i}v_j(\bm{x}\times\Sp)_{k>} \nn \\
&& + \big(2 - 6\eta\big)x_{<i}v_j(\bm{x}\times\D)_{k>}\Bigg\},
\eea
\bea
I_{ijkl} &=& \eta m\Bigg\{\Bigg[1 - 3\eta + \frac{1}{c^2}\bigg[\left(\frac{103}{110} - \frac{147}{22}\eta + \frac{279}{22}\eta^2\right)v^2 + \left(-\frac{10}{11} + \frac{61}{11}\eta - \frac{105}{11}\eta^2\right)\mr \bigg]\Bigg]x_{<ijkl>} \nn \\
&& - \frac{72}{55}\frac{r\rd}{c^2}\big(1 - 5\eta + 5\eta^2\big)v_{<i}x_{jkl>} + \frac{78}{55}\frac{r^2}{c^2}\big(1-5\eta + 5\eta^2\big)v_{<ij}x_{kl>}\Bigg\},
\eea
\bea
I_{ijklm} &=& -\eta \, \delta m(1-2\eta)x_{<ijklm>},
\eea
\bea
J_{ij} &=& \eta \,\delta m \Bigg\{\Bigg[-1 + \frac{1}{c^2}\bigg[\left(-\frac{13}{28} + \frac{17}{7}\eta\right)v^2 + \left(-\frac{27}{14} - \frac{15}{7}\eta\right)\mr\bigg] + \frac{1}{c^4}\bigg[\left(-\frac{29}{84} +\frac{11}{3}\eta - \frac{505}{56}\eta^2\right)v^4 \nn \\
&& + \left(-\frac{671}{252} + \frac{1297}{126}\eta + \frac{121}{12}\eta^2\right)\mr v^2 + \left(\frac{5}{252} + \frac{241}{252}\eta + \frac{335}{84}\eta^2\right)\mr \rd^2 \nn \\
&& + \left(\frac{43}{252} + \frac{1543}{126}\eta - \frac{293}{84}\eta\right)\mrsq\bigg]\Bigg](\bm{x}\times\bv)_{<i}x_{j>} + \frac{r\rd}{c^2}\Bigg[-\frac{5}{28} + \frac{5}{14}\eta + \frac{1}{c^2}\bigg[\left(-\frac{25}{168} + \frac{25}{24}\eta - \frac{25}{14}\eta^2\right)v^2 \nn \\
&& + \left(-\frac{103}{63} - \frac{337}{126}\eta + \frac{173}{84}\eta^2\right)\mr\bigg]\Bigg](\bm{x}\times\bv)_{<i}v_{j>}\Bigg\} +  \frac{\eta}{c}\Bigg\{-\frac{3}{2}x_{<i}\Delta_{j>} + \frac{1}{c^2}\Bigg[r\rd\left(\frac{3}{7} - \frac{16}{7}\eta\right)v_{<i}\Delta_{j>}\nn \\
&&  + r\rd\frac{3}{7}\dm v_{<i}S_{j>} + \bigg[\left(\frac{27}{14} - \frac{109}{14}\eta\right)(\bv\cdot\D)  + \frac{27}{14}\dm(\bv\cdot\Sp)\bigg]_{x_<i}v_{j>} \nn \\
&& + \bigg[\left(-\frac{11}{14} + \frac{47}{14}\eta\right)(\bm{x}\cdot\D) - \frac{11}{14}\dm(\bm{x}\cdot\Sp)\bigg]v_{<ij>} + \bigg[\left(-\frac{29}{28} + \frac{143}{28}\eta\right)v^2 + \left(\frac{19}{28} + \frac{13}{28}\eta\right)\mr\bigg]x_{<i}\Delta_{j>} \nn \\
&& + \frac{1}{r^2}\mr\bigg[\left(-\frac{4}{7} + \frac{31}{14}\eta\right)(\bm{x}\cdot\D) - \frac{29}{14}\dm(\bm{x}\cdot\Sp)\bigg]x_{<ij>} + \dm\bigg[-\frac{2}{7}v^2 - \frac{1}{14}\mr\bigg]x_{<i}S_{j>}\Bigg]\Bigg\}
\eea
\bea
J_{ijk} &=& \eta m\Bigg\{\Bigg[1 - 3\eta + \frac{1}{c^2}\bigg[\left(\frac{41}{90} - \frac{77}{18}\eta + \frac{185}{18}\eta^2\right)v^2 + \left(\frac{14}{9} - \frac{16}{9}\eta - \frac{86}{9}\eta^2\right)\mr\bigg]\Bigg](\bm{x}\times\bv)_{<i}x_{jk>} \nn \\
&& + \frac{7}{45}\frac{r^2}{c^2}\big(1 - 5\eta + 5\eta^2\big)(\bm{x}\times\bv)_{<i}v_{jk>} + \frac{2}{9}\frac{r\rd}{c^2}\big(1 - 5\eta + 5\eta^2\big)(\bm{x}\times\bv)_{<i}x_jv_{k>}\Bigg\} \nn \\
&& + \frac{2\eta}{c}\Bigg\{x_{<ij}S_{k>} + \dm x_{<ij}\Delta_{k>}  \Bigg\},
\eea
\bea
J_{ijkl} &=& -\eta \delta m \,(1-2\eta)(\bm{x}\times\bv)_{<i}x_{jkl>}.
\eea \end{widetext}
It is important to note here that in principle there should also be a contribution in $I_{ij}$ at 2PN order from the individual quadrupole moment of each black hole. Indeed by dimensional analysis one finds that the mass quadrupole of a Kerr black hole scales as, restoring factors of $G$ and $c$ for clarity here,
\be
Q_{\rm Kerr} \sim \frac{G}{c^2} \frac{1}{m}S_{\rm true}^2 \sim \frac{1}{c^4} \frac{1}{m} S^2,
\ee
where we used the PN scaling between the true (physical) spin and the spin variable with finite limit as $c\rightarrow \infty$ discussed in the introduction. Since the first time derivative of this quadrupole moment comes from the spin precession equation only, which is ${\cal O}(2)$, the contribution to the linear momentum flux from time derivatives of the individual Kerr quadrupoles is pushed to 3PN order, and can thus be ignored here. This completes the list of all required source multipole moments for the computation of the linear momentum flux at 2PN order for spinning binaries.

\section{Construction of quasi-circular orbits}\label{CircRedux}

When spins are present, exact circular motion in a fixed orbital plane is not a solution to equations of motion~\eqref{accs}. The spins induce radial and azimuthal perturbations, as well as precession of the orbital plane. In this Appendix we provide, for the benefit of the unfamiliar reader, a review of the well-known fact that despite these difficulties, it is still possible to meaningfully derive a modified version of Kepler's law for spinning binaries. This modified Kepler's law relates the orbit-averaged orbital frequency and the orbit-averaged orbital separation, as given in Eq.\eqref{modKepler}. 

Our description of the orbit follows exactly the formalism of Ref.~\cite{ABFO}. The basic picture is the following. One describes the orbit using the unit vector $\hn$ along the line joining the two bodies, the unit vector $\hLN$ normal to the instantaneous orbital plane, and the vector $\hbv = \hLN \times \hn$. With respect to this basis, the instantaneous velocity and acceleration are shown to be
\bes
\bea
\bv &=& \rd \hn + r\omega \hbv , \\
\bm{a} &=& (\ddot{r} - r\omega^2)\hn + (r\dot{\omega} + 2\rd \omega)\hbv - r\omega\left(\hbv\cdot\frac{d\hLN}{dt}\right)\hLN \nn \\
&=& \bm{a}_{\mathrm{N}}(r,\hn) + \frac{1}{c^2}\bm{a}_{\mathrm{1PN}}(r,\rd,\hn,\bv) +  \frac{1}{c^4}\bm{a}_{\mathrm{2PN}}(r,\rd,\hn,\bv) \nn \\ && +  \frac{1}{c^3}\bm{a}_{\mathrm{SO}}(r,\rd,\hn,\bv,\Sp,\D) +  \frac{1}{c^4}\bm{a}_{\mathrm{SS}}(r,\hn,\Sp_0). \label{accs2}
\eea
\ees
Our quasi-circular orbits are then constructed as follows. Since the leading-order spin acceleration is of 1.5PN order, we assume that the radial perturbations scale similarly, i.e. $\rd \sim O(3)$. Hence at 2PN accuracy we may set $\rd = 0$ and $\bv = r\omega\hbv$ in the arguments of each acceleration term in Eq.~\eqref{accs2}. By projecting the result on the $(\hn,\hbv,\hLN)$ triad, we find (setting $c=1$)
\bea
\bm{a} &=& -\frac{m}{r^2}\Bigg[1 + (1+3\eta)v^2 - 2(2+\eta)\mr + \eta(3-4\eta)v^4 \nn \\ &&  - \frac{1}{2}\eta(13-4\eta)\mr v^2 - \frac{\omega}{m} \hLN\cdot\bigg(5\Sp + 3\dm \D\bigg) \nn \\
&& + \frac{3}{4}(12+29\eta)\mrsq + \frac{3}{2m^2r^2}\Big(\Sp_0^2 - 3(\hn\cdot\Sp_0)^2\Big)\Bigg]\hn \nn \\ &&  - \frac{3}{mr^4}(\hn\cdot\Sp_0)(\hbv\cdot\Sp_0)\hbv  + \left[\frac{\omega}{r^2}\hn\cdot\bigg(7\Sp + 3\dm \D\bigg) \right. \nn \\ && \left.- \frac{3}{mr^4}(\hn\cdot\Sp_0)(\hLN\cdot\Sp_0)\right]\hLN,    
\eea
where $v^2 = r^2\omega^2$. Next we decompose $r$ and $\omega$ into their orbital average piece plus a time-dependent fluctuation, i.e.
\bes\label{OrbDecomp}
\bea
r &=& \bar{r} + \delta r ,\\
\omega &=& \bar{\omega} + \delta \omega.
\eea
\ees
The radial motion and the (time dependent) orbital frequency are determined from the $\hn$ and $\hbv$ components of the equation of motion~\eqref{accs2}. Let us first look at the component of the equation of motion along $\hbv$. It yields
\be
r\dot{\omega} + 2\rd \omega = \frac{1}{r}\frac{d}{dt}(r^2 \omega) = -\frac{3}{mr^4} (\hn\cdot\Sp_0)(\hbv\cdot\Sp_0). \label{EOMlambda}
\ee
In order to perform the integral, we may keep the spins constant, as their time derivatives yield higher order terms. We may also use $\hbv = \omega^{-1} \dot{\hn}$ in the right-hand side of Eq.~\eqref{EOMlambda}, which then gives
\bea
\frac{d}{dt}(r^2\omega) &=& -\frac{3}{mr^3\omega}(\hn\cdot\Sp_0)\left(\frac{d\hn}{dt}\cdot\Sp_0\right) \nn \\  &=&  -\frac{3}{2mr^3\omega}\frac{d}{dt}(\hn\cdot\Sp_0)^2.
\eea
On the right-hand side we may assume that $r$ and $\omega$ are constants, as their time derivatives are at least $O(3)$. Hence the 2PN accurate solution to the $\hbv$ component of the equations of motion yields
\be\label{AngMomEq}
r^2\omega = -\frac{3}{2mr^3\omega}(\hn\cdot\Sp_0)^2 + \kappa,
\ee
where $\kappa$ is an integration constant. Substituting decomposition~\eqref{OrbDecomp} into Eq.~\eqref{AngMomEq} we find
\be\label{AngMomEq2}
\bar{r}^2 \bar{\omega} + 2\bar{r}\bar{\omega}\,\delta r + \bar{r}^2 \delta \omega = -\frac{3}{2m\bar{r}^3\bar{\omega}}(\hn \cdot\Sp_0)^2 + \kappa.
\ee
Since, by definition, $\delta r$ and $\delta \omega$ have zero orbital average, the constant $\kappa$ is determined as 
\be
\kappa = \bar{r}^2 \bar{\omega} + \frac{3}{4m\bar{r}^3\bar{\omega}}\big[\Sp_0^2 - (\hLN\cdot\Sp_0)^2\big],
\ee
where we have used the following orbital average
\be
\langle \hn^i \hn^j \rangle = \frac{1}{2}(\delta^{ij} - \hLN^i \hLN^j) + \mathcal{O}(3).
\ee
Equation~\eqref{AngMomEq2} becomes
\be\label{AngMomEq3}
2\bar{r}\bar{\omega}\,\delta r + \bar{r}^2 \delta \omega = \frac{3}{4m\bar{r}^3\bar{\omega}}\big[\Sp_0^2 - (\hLN\cdot\Sp_0)^2\big] -\frac{3}{2m\bar{r}^3\bar{\omega}}(\hn \cdot\Sp_0)^2.
\ee
Let us now look at the $\hn$ component of the equation of motion. It yields
\bea
\delta\ddot{r} &=& \bar{r}\bar{\omega}^2 + \bar{\omega}^2 \delta r + 2\bar{r}\bar{\omega}\,\delta{\omega} 
-\frac{m}{\bar{r}^2}\Bigg(1 - 2\frac{\delta r}{\bar{r}}\Bigg)\Bigg[1 + \nn \\ && (1+3\eta)\bar{r}^2\bar{\omega}^2 - 2(2+\eta)\frac{m}{\bar{r}} + \eta(3-4\eta)\bar{r}^4\bar{\omega}^4 \nn \\
&& - \frac{1}{2}\eta(13-4\eta)\bar{r}^2\bar{\omega}^2\frac{m}{\bar{r}} + \frac{3}{4}(12+29\eta)\frac{m^2}{\bar{r}^2} \nn \\
&& - \frac{\bar{\omega}}{m} \hLN\cdot\bigg(5\Sp + 3\dm \D\bigg) \nn \\ && + \frac{3}{2m^2\bar{r}^2}\Big(\Sp_0^2 - 3(\hn\cdot\Sp_0)^2\Big)\Bigg] \nn \\
&\equiv& 2\frac{m}{\bar{r}^3}\delta r - \frac{m}{\bar{r}^2}W + \frac{9}{2m\bar{r}^4}(\hn \cdot\Sp_0)^2,  \label{modKeplerIntermediate}
\eea
where $W$ can be taken as constant for the purpose of solving Eq.~\eqref{modKeplerIntermediate} at 2PN order. Taking the orbital average of Eq.~\eqref{modKeplerIntermediate} we find, imposing $\langle \delta \ddot{r} \rangle = 0$
\bea
\bar{r}^2 \bar{\omega}^2 &=& \frac{m}{\bar{r}}W - \frac{9}{4m\bar{r}^3}\big[\Sp_0^2 - (\hLN\cdot\Sp_0)^2\big] \nn \\
&=& \frac{m}{\bar{r}}\Bigg[1 + (1+3\eta)\bar{r}^2\bar{\omega}^2 - 2(2+\eta)\frac{m}{\bar{r}} \nn \\ && + \eta(3-4\eta)\bar{r}^4\bar{\omega}^4  - \frac{1}{2}\eta(13-4\eta)\bar{r}^2\bar{\omega}^2\frac{m}{\bar{r}} \nn \\ && + \frac{3}{4}(12+29\eta)\frac{m^2}{\bar{r}^2} - \frac{\bar{\omega}}{m} \hLN\cdot\bigg(5\Sp + 3\dm \D\bigg)\nn \\ &&  - \frac{3}{4m^2\bar{r}^2}\Big(\Sp_0^2 - 3(\hLN\cdot\Sp_0)^2\Big) \Bigg].\label{modKeplerIntermediate2}
\eea
Solving Eq.~\eqref{modKeplerIntermediate2} for $\bar{\omega}$ at 2PN accuracy, we recover Eq.~\eqref{modKepler} after performing the replacements $\bar{r} \rightarrow r$ and $\bar{\omega} \rightarrow \omega$. It remains to solve for the radial perturbation. Substituting Eqs.~\eqref{AngMomEq3} and~\eqref{modKeplerIntermediate2} into Eq.~\eqref{modKeplerIntermediate} we obtain the following evolution equation for $\delta r$
\be
\delta\ddot{r} + \bar{\omega}^2 \delta r = \frac{3}{2m\bar{r}^4}\Bigg[(\hn\cdot\Sp_0)^2 - \frac{1}{2}\big[\Sp_0^2 - (\hLN\cdot\Sp_0)^2\big] \Bigg].
\ee 
The 2PN accurate general solution to this differential equation is
\bea
\delta r &=& A \cos(\bar{\omega} t + \varphi) + \frac{1}{2 m^2\bar{r}}(\hn \cdot\Sp_0)^2 + \frac{1}{m^2\bar{r}}(\hbv\cdot\Sp_0)^2 \nn \\ && - \frac{3}{4m^2\bar{r}}\big[\Sp_0^2 - (\hLN\cdot\Sp_0)^2\big],
\eea
where $A$ and $\varphi$ are constants determined by initial conditions. For definiteness we assume $A=0$, so that the homogeneous solution to the radial perturbation vanishes. The last element we need is the angular frequency perturbation, which is given by
\bea
\delta \omega &=& \frac{\bar{\omega}}{m^2\bar{r}^2}\Bigg\{-\frac{5}{2}(\hn\cdot\Sp_0)^2 - 2(\hbv \cdot\Sp_0)^2 \nn \\ && + \frac{9}{4}\big[\Sp_0^2 - (\hLN \cdot\Sp_0)^2\big]\Bigg\}.
\eea
The perturbations to the orbital frequency and radial motion are quadratic in the spins and therefore are 2PN corrections. Hence we only need to make the distinction between $r$ and $\bar{r}$ and $\omega$ and $\bar{\omega}$ in the $\mathcal{O}(0)$ (Newtonian) piece of linear momentum flux.

%Substituting Eq.~\eqref{AngMomEq} into the $\hn$ component of the equation of motion, we find
%\bea
%\ddot{r} - r\omega_{\mathrm{orb}}^2 &=& \ddot{r} - \kappa \frac{\omega_{\mathrm{orb}}}{r} + \frac{3}{2mr^4}(\hn\cdot\Sp_0)^2 \nn \\
%&=& -\frac{m}{r^2}\Bigg[1 + (1+3\eta)v^2 - 2(2+\eta)\mr + \eta(3-4\eta)v^4 - \frac{1}{2}\eta(13-4\eta)\mr v^2 + \frac{3}{4}(12+29\eta)\mrsq \nn \\
%&& - \frac{\omega_{\mathrm{orb}}}{m} \hLN\cdot\bigg(5\Sp + 3\dm \D\bigg) + \frac{3}{2m^2r^2}\Big(\Sp_0^2 - 3(\hn\cdot\Sp_0)^2\Big)\Bigg].
%\eea 
%Inside the bracket, since the time derivatives of $r$ and $\omega_{\mathrm{orb}}$ are $O(3)$, we may consider them as constants.
\section{Computation of tail integral}\label{IntTail}

We provide here, for completeness and also for the unfamiliar reader,
an explicit derivation of Eq.~\eqref{keyint}, central to the
evaluation of tail terms in the limit of quasi-circular
orbits. 

\begin{figure}[here]
\includegraphics[height=6cm]{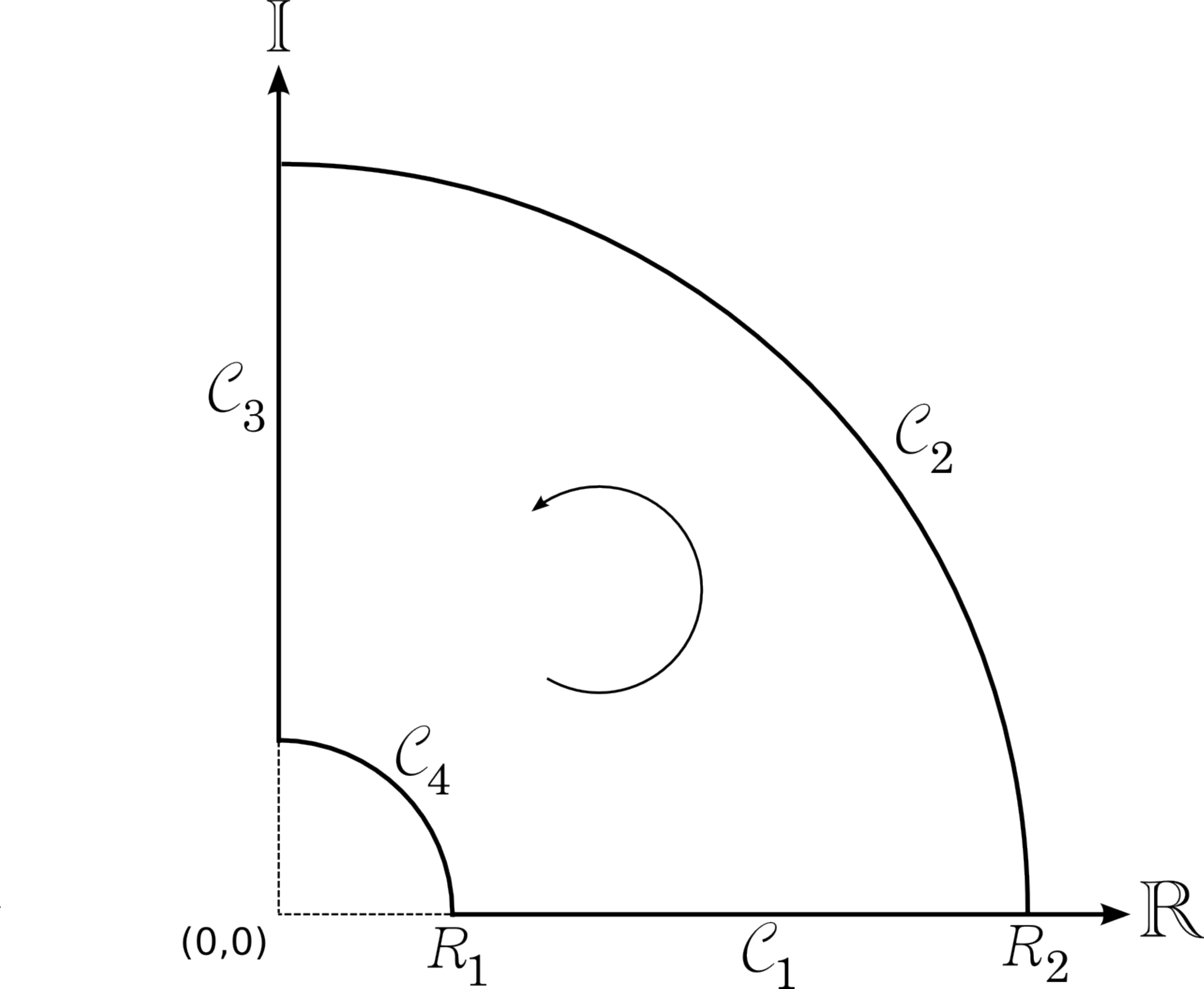}
\caption{Contour of integration for evaluating the tail integral.}\label{fig:contour}
\end{figure}

Consider first the integral 
\be\label{int1} \int_0^\infty \,\ln x \, e^{ikx} \, dx, 
\ee 
which can be mapped directly to Eq.~\eqref{keyint} by a simple change of 
integration variable and a redefinition of $k$. To evaluate Eq.~\eqref{int1}, 
let us examine the following contour integral 
\be
\label{contourint} \oint_{\cal C} \, \ln z \,e^{ikz} \, dz, 
\ee 
where the contour ${\cal C}$ is the closed contour obtained 
by the union of the paths ${\cal C}_1$, ${\cal C}_2$, 
${\cal C}_3$ and ${\cal C}_4$ taken counter-clockwise as depicted in Fig.~\ref{fig:contour}.

On the contour ${\cal C}$ and inside the region it borders, we take $\ln z$
to be the principal logarithm $\ln z = \ln r + i\theta$, with $z =
r e^{i\theta}$. The principal logarithm is analytic on and inside the contour 
${\cal C}$, and therefore Eq.~\eqref{contourint} vanishes. Let us now go through each
path which makes up the contour of integration. First, the integral from $R_1$ to $R_2$ 
on the real axis (the path ${\cal C}_1$ in Fig.~\ref{fig:contour}) 
matches Eq.~\eqref{int1} when the limits $R_1\rightarrow 0$ and 
$R_2\rightarrow \infty$ are taken. [This, of course,
is the reason why the principal branch of $\ln z$ is chosen to perform
the integral~\eqref{contourint}.]  Next we have the contribution from the
path ${\cal C}_2$, i.e., the quarter-circle of radius $R_2$ in Fig.~\ref{fig:contour}. 
This is given by 
\bea
\int_{{\cal C}_2} \, \ln z \, e^{ikz} \, dz &=& iR_2 \int_0^{\pi/2} (\ln R_2 +
i\theta) e^{ikR_2\cos\theta} \times \nn \\ && e^{-kR_2\sin\theta} e^{i\theta} d\theta\,.\label{intR} 
\eea 
Clearly this integral vanishes in the limit $R_2 \rightarrow
\infty$ due to the presence of the $e^{-kR_2\sin\theta}$ factor in the
integrand. The integral of the path ${\cal C}_4$, i.e., the 
quarter-circle of radius $R_1$ is quite similiar to Eq.~\eqref{intR}, 
and is given by 
\bea
\int_{{\cal C}_4} \, \ln z \, e^{ikz} \, dz &=& iR_1 \int_{\pi/2}^{0} (\ln R_1 +
i\theta) e^{ikR_1\cos\theta} \times \nn \\ && e^{-kR_1\sin\theta} e^{i\theta} d\theta. \label{intr} 
\eea
Since $R_1\ln R_1 \rightarrow 0$ as $R_1 \rightarrow 0$, Eq.\eqref{intr}
clearly vanishes in the limit $R_1 \rightarrow 0$. Lastly the integral
over the path ${\cal C}_3$ , i.e., the positive imaginary axis, is 
\bea
\int_{{\cal C}_3} \ln z \, e^{ikz} \, dz &=& \int_{R_2}^{R_1} \left (\ln y + i\frac{\pi}{2}\right) e^{-ky} \, (idy), \nn \\
&=& \frac{\pi}{2k}\left (e^{-kR_1} - e^{-kR_2}\right ) \nn \\ && - i\int_{R_1}^{R_2} \ln y \,
e^{-ky} \, dy.  
\label{eq:C3}
\eea 
Integrating by parts the second term in the right hand side of Eq.~(\ref{eq:C3}) yields
\bea
\int_{R_1}^{R_2} \ln y \, e^{-ky} \, dy &=& -\frac{1}{k}\left[\ln y \, e^{-ky}\right]_{R_1}^{R_2} + \frac{1}{k}\int_{R_1}^{R_2} 
e^{-ky} \frac{dy}{y} \nn \\
&=&-\frac{1}{k} \ln R_2 \, e^{-kR_2} + \frac{1}{k}\ln (R_1)\, e^{-kR_1} \nn \\ && + \frac{1}{k}\int_{k R_1}^{R_2} e^{-u} \frac{du}{u} .
\eea 
We next make use of the following integral representation for the logarithm of a number approaching zero 
\be
\int_\alpha^\infty -\frac{e^{-u}}{1-e^{-u}}du = \ln \big[1 - e^{-\alpha}\big] \rightarrow \ln \alpha \quad {\rm as} \quad 
\alpha \rightarrow 0. 
\ee
Thus, taking the limits $R_1 \rightarrow 0$ and
$R_2\rightarrow\infty$ we obtain 
\be \int_0^\infty k \ln y \, e^{-ky} \,dy = -\ln k + \int_0^\infty
\left(\frac{e^{-u}}{u} - \frac{e^{-u}}{1-e^{-u}}\right)du.  
\ee 
The remaining integral can be recognized as the digamma function $\Psi(x)
= d \ln \Gamma(x) / dx$ evaluated at $x=1$. It is well known that
$\Psi(1) = -\gamma_{\rm E}$, $\gamma_{\rm E}$ being the Euler-Mascheroni constant. 
Thus, the final result is
\bea
\int_0^\infty \ln x \, e^{ikx}\, dx &=& - \int_\infty^0  \left[\ln y + i\frac{\pi}{2}\right] e^{-ky} \, (idy) \nn \\
&=& -\frac{\pi}{2k} - i\left[\frac{1}{k}\ln k + \gamma_{\rm E}
\right]. \label{keyint2} 
\eea 
Setting $x = u / 2B$ and $k \equiv 2Bn\omega$ in Eq.~\eqref{keyint2}, we recover Eq.~\eqref{keyint}.


\begin{thebibliography}{0}

\bibitem{Merritt}
D. Merritt {\it et al.},  Astrophys. J. {\bf 607}, L9 (2004).

\bibitem{MQ}
P. Madau and E. Quataert,  Astrophys. J. {\bf 606}, L17 (2004).

\bibitem{Haiman}
Z. Haiman,  Astrophys. J. {\bf 613}, 36 (2004).

\bibitem{TH}
T. Tanaka and Z. Haiman, arXiv:0807.4702.

\bibitem{VP}
M. Volonteri and R. Perna, Mon. Not. Roy. Astron. Soc. {\bf 358}, 913 (2005).

\bibitem{Volonteri}
M. Volonteri,  Astrophys. J. {\bf 663}, L5 (2007).

\bibitem{BL}
L. Blecha and A. Loeb, arXiv:0805.1420.

\bibitem{LCFH}
N. I. Libeskind, S. Cole, C. S. Frenk and J. C. Helly, Mon. Not. Roy. Astron. Soc. {\bf 368}, 1381 (2006).

\bibitem{MAS}
M. Micic, T. Abel and S. Sigurdsson, Mon. Not. Roy. Astron. Soc. {\bf 372}, 1540 (2006).

\bibitem{BKMQ}
M. Boylan-Kolchin, C.-P. Ma and E. Quataert,  Astrophys. J. {\bf 613}, L37 (2004).

\bibitem{GM}
A. Gualandris and D. Merritt,  Astrophys. J. {\bf 678}, 780 (2008).

%\bibitem{HDR}
%M. G. Haehnelt, M. B. Davies and M. J. Rees, Mon. Not. Roy. Astron. Soc. {\bf 366}, L22 (2005).

\bibitem{KZL}
S. Komossa, H. Zhou and H. Lu,  Astrophys. J. {\bf 678}, L81 (2008).

\bibitem{Bogdanovic}
T. Bogdanovic, M. Eracleous and S. Sigurdsson, arXiv:0809.3262.

\bibitem{Dotti}
M. Dotti, C. Montuori, R. Decarli, M. Volonteri, M. Colpi and F. Haardt, arXiv:0809.3446.

%\bibitem{Loeb}
%A. Loeb, Phys. Rev. Lett. {\bf 99}, 041103 (2007).

\bibitem{Baker2}
J. G. Baker, J. Centrella, D.-I. Choi, M. Koppitz, J. R. van Meter and M. C. Miller, Astrophys. J. {\bf 653}, L93 (2006).

\bibitem{Pollney}
D. Pollney, C. Reisswig, L. Rezzolla, B. Szilagyi, M. Ansorg, B. Deris, P. Diener, E. N. Dorband, M. Koppitz, A. Nagar and E. Schnetter, Phys. Rev D {\bf 76}, 124002 (2007).

\bibitem{Koppitz}
M. Koppitz {\it et al.}, Phys. Rev. Lett. {\bf 99}, 041102 (2007).

\bibitem{jena-no-spin} J. A. Gonz\'{a}lez, U. Sperhake, B. Br\"{u}gmann, M. Hannam, and S. Husa, Phys. Rev. Lett. \textbf{98}, 091101
  (2007). 

\bibitem{Herrmann}
F. Herrmann, I. Hinder, D. M. Shoemaker, P. Laguna and R. A. Matzner, Phys. Rev. D {\bf 76}, 084032 (2007); {\it ibid.},  Astrophys. J. {\bf 661}, 430 (2007).

\bibitem{Gonzalez} J.A. Gonzalez, M. Hannam, U. Sperhake, B. Bruegmann, and S. Husa, Phys. Rev. Lett. {\bf 98}, 231101 (2007).

\bibitem{Campanelli}
M. Campanelli, C. O. Lousto, Y. Zlochower and D. Merritt, Phys. Rev. Lett {\bf 98} 231102 (2007); {\it ibid.}, Astrophys. J. {\bf 659}, L5 (2007).

\bibitem{recoilFAU} W. Tichy and P. Marronetti, Phys. Rev. D {\bf 76}, 061502 (2007).

\bibitem{Bruegmann}
B. Br\"{u}gmann, J. A. Gonzalez, M.Hannam, S. Husa and U. Sperhake, Phys. Rev. D {\bf 77}, 124047 (2008).

\bibitem{Baker}
J. G. Baker, W. D. Boggs, J. Centrella, B. J. Kelly, S. T. McWilliams, M. C. Miller and J. R. van Meter, Astrophys. J. {\bf 682}, L29 (2008).

\bibitem{Lousto1}
C. O. Lousto and Y. Zlochower, Phys. Rev. D {\bf 77}, 044028 (2008).

\bibitem{Lousto2}
S. Dain, C. O. Lousto and Y. Zlochower, arXiv:0803.0351.

\bibitem{Lousto3}
C. O. Lousto and Y. Zlochower, arXiv:0805.0159.

\bibitem{Fitchett1}
M. Fitchett, Mon. Not. Roy. Astron. Soc. {\bf 203}, 1049 (1983).

\bibitem{Fitchett2}
M. Fitchett and S. Detweiler, Mon. Not. Roy. Astron. Soc. {\bf 211}, 933 (1984).

\bibitem{Peres}
A. Peres, Phys. Rev. {\bf 128}, 2471 (1962).

\bibitem{Bekenstein}
J. D. Bekenstein, Astrophys. J. {\bf 183}, 657 (1973).

\bibitem{Wiseman}
A. G. Wiseman, Phys. Rev. D {\bf 46}, 1517 (1992).

\bibitem{BQW}
L. Blanchet, M. S. S. Qusailah and C. M. Will, Astrophys. J. {\bf 635}, 508 (2005).

\bibitem{Kidder}
L. E. Kidder, Phys. Rev. D {\bf 52}, 821  (1995).

\bibitem{Racine}
\'{E}. Racine, Phys. Rev. D {\bf 78}, 044021 (2008).

\bibitem{DamourSpinEOB}
T. Damour, Phys. Rev. D {\bf 64}, 124013 (2001).

\bibitem{DG}
T. Damour and A. Gopakumar, Phys. Rev. D {\bf 73}, 124006 (2006).

\bibitem{anatomy} J.D. Schnittman, A. Buonanno, J. R. van Meter, J. G. Baker, W. D. Boggs, 
J. Centrella, B. J. Kelly, and S. T. McWilliams, Phys. Rev. D {\bf 77}, 044031 (2008). 

\bibitem{SB} J. Schnittman and A. Buonanno, Astrophys. J. {\bf 662}, L63 (2007).

\bibitem{EOB1}
A. Buonanno and T. Damour, Phys. Rev. D {\bf 59}, 084006 (1999).

\bibitem{EOB2}
A. Buonanno and T. Damour, Phys. Rev. D {\bf 62}, 064015 (2000).

\bibitem{EOB3}
T. Damour, P. Jaranowski and G. Sch\"{a}fer, Phys. Rev. D {\bf 62}, 084011 (2000).

\bibitem{Tagoshi}
H. Tagoshi, M. Shibata, T. Tanaka and M. Sasaki, Phys.Rev. D {\bf 54}, 1439  (1996).

\bibitem{MVG}
B. Mik\'{o}czi, M. Vas\'{u}th and L. \'{A}. Gergely, Phys. Rev. D {\bf 71}, 124043 (2005).

\bibitem{HNLSO}
T. Damour, P. Jaranowski and G. Sch\"{a}fer, Phys. Rev. D {\bf 77}, 064032 (2008).

\bibitem{Porto}
R. A. Porto, Phys. Rev. D {\bf 73}, 104031 (2006).

\bibitem{FBB}
G. Faye, L. Blanchet and A. Buonanno, Phys. Rev. D {\bf 74}, 104033 (2006).

\bibitem{WS}
C. M. Will and A. G. Wiseman, Phys. Rev. D {\bf 54}, 4813 (1996).

\bibitem{PortoS1S2}
R. A. Porto and I. Z. Rothstein, Phys. Rev. D {\bf 78}, 044012 (2008).

\bibitem{PortoS1S1}
R. A. Porto and I. Z. Rothstein, Phys. Rev. D {\bf 78}, 044013 (2008).

\bibitem{SpinSchafer1}
J. Steinhoff, G. Sch\"{a}fer and S. Hergt, Phys. Rev. D {\bf 77}, 104018 (2008).

\bibitem{SpinSchafer2}
J. Steinhoff, S. Hergt and G. Sch\"{a}fer, Phys. Rev. D {\bf 78}, 101503(R) (2008).

\bibitem{SpinSchafer3}
S. Hergt and G. Sch\"{a}fer, arXiv:0809.2208 (gr-qc).

\bibitem{Wald}
R. Wald, Phys. Rev. D {\bf 6}, 406 (1972).

\bibitem{Beiglbock}
W. Beiglb\"{o}ck, Commun. Math. Phys. {\bf 5}, 112 (1964).

\bibitem{Madore}
J. Madore, Ann. Inst. Henri Poincar\'{e} {\bf A11}, 221 (1969).

\bibitem{Dixon}
W. G. Dixon, Proc. Roy. Soc. Lond. {\bf A314}, 499 (1970).

\bibitem{Tulczyjew}
W. Tulczyjew, Acta Phys. Pol. {\bf 18}, 393 (1959).

\bibitem{Pirani}
F. A. E. Pirani, Acta Phys. Pol. {\bf 15}, 389 (1956).

\bibitem{MTW}
C. W. Misner, K. S. Thorne and J. A. Wheeler, {\it Gravitation}, W. H. Freeman and Company, New York (1973).
\bibitem{KyrianSemerak}
K. Kyrian and O. Semer\'{a}k, Mon. Not. R. Astron. Soc. {\bf 382}, 1922 (2007).

\bibitem{Papapetrou}
A. Papapetrou, Proc. R. Soc. Lond. {\bf A209}, 248 (1951).

\bibitem{Weinberg}
S. Weinberg, {\it Gravitation and Cosmology: principles and applications of the general theory of relativity}, John Wiley \& Sons Inc., 1972.

\bibitem{Owen}
H. Tagoshi, A. Ohashi and B. J. Owen, Phys. Rev. D {\bf 63}, 044006 (2001).

\bibitem{BOC}
B. M. Barker and R. F. O'Connell, Gen. Relativ. Gravit. {\bf 11}, 149 (1979).

\bibitem{Poisson}
\'{E}. Poisson, Phys. Rev. D {\bf 57}, 5287 (1998).

\bibitem{Gergely}
L. \'{A}. Gergely and Z. Keresztes, Phys. Rev. D {\bf 67}, 024020 (2003).

\bibitem{Racine2}
\'{E}. Racine, Class. Quant. Grav. {\bf 23}, 373 (2006).

\bibitem{BBF}
L. Blanchet, A. Buonanno and G. Faye, Phys. Rev. D {\bf 74}, 104034 (2006); Erratum-ibid. D {\bf 75}, 049903 (2007).

\bibitem{LB} L. Blanchet, Living Rev. Rel. {\bf 5}, 3 (2002).

\bibitem{Thorne}
K. S. Thorne, Rev. Mod. Phys {\bf 52}, 299 (1980).

\bibitem{BlanchetDamour} L. Blanchet and T. Damour, Phys. Rev. D {\bf 46}, 4304 (1992).

\bibitem{BlanchetSchafer}
L. Blanchet and G. Sch\"afer, Class. Quant. Grav. {\bf 10}, 2699 (1993).

\bibitem{SpinExpansion}
L. Boyle, M. Kesden and S. Nissanke, Phys. Rev. Lett. {\bf 100}, 151101 (2008).

\bibitem{FP} F. Pretorius, {\it Binary Black Hole Coalescence}, arxiv:0710.1338 (gr-qc).

\bibitem{Rieth}
R. Rieth and G. Sch\"{a}fer, Class. Quant. Grav. {\bf 14}, 2357 (1997).

\bibitem{GI}
A. Gopakumar and B. R. Iyer, Phys. Rev. D {\bf 56}, 7708 (1997).

\bibitem{Boyleetal08} M. Boyle, A. Buonanno, L. Kidder, A. Mroue, Y. Pan, H. Pfeiffer and M.A. Scheel, 
arXiv:0804.4184 (gr-qc).

\bibitem{ABFO}
K. G. Arun, A. Buonanno, G. Faye and E. Ochsner, arxiv:0810.5336 (gr-qc).




\end{thebibliography}
\end{document}